\documentstyle[12pt]{article}
\newcommand{\gsim}{\:\raisebox{.25ex}{$>$}\hspace*{-.75em}
      \raisebox{-.93ex}{$\sim$}\:}
\newcommand{\lsim}{\:\raisebox{.25ex}{$<$}\hspace*{-.75em}
      \raisebox{-.93ex}{$\sim$}\:}

\newcommand{\be}{\begin{equation}}
\newcommand{\ee}{\end{equation}}

\begin{document}

\begin{titlepage}
\hfill                                              Preprint ITEP-61(1994)
\vspace{3mm}

\centerline{\large\bf THE ~NUCLEON ~SPIN ~PROBLEM}
\vspace{30mm}

\centerline{\bf B.L.IOFFE}
\vspace{10mm}

\centerline{\it Institute of Theoretical and Experimental Physics,}
\vspace{3mm}

\centerline{\it B.Cheremushkinskaya 25, 117259, Moscow, Russia}
\vspace{30mm}

\noindent
{\bf C o n t e n t s}
\vspace{5mm}

\noindent
1. ~Introduction.\\
2. ~General Relations.\\
3. ~Small $x$ Domain -- Regge Behaviour.\\
4. ~The Parton Model.\\
5. ~Sum Rules in the Parton Model.\\
6. ~QCD Corrections to Sum Rules.\\
7. ~Nucleon Axial Coupling Constants.\\
8. ~The Experimental Data on $g_1(x)$ and Their Interpretation.\\
9. ~Suggestions for Future Experiments.\\
10. The Structure Function $g_2(x)$.\\
11. Chirality Violating Structure Function $h_1(x)$.\\
12. Conclusions.
\vspace{3mm}

{\it Lectures, presented at ITEP Winter School, March 1994}
\vspace{10mm}
\end{titlepage}

\newpage

\noindent
{\large\bf 1. Introduction}
\vspace{3mm}

For many years there exists a permanent non passing away interest to the
nucleon spin problem: how the nucleon spin is distributed among its
constituents -- quarks and gluons. This interest arose essentially after
appearance of the results of famous the EMC experiment $^1$ on deep
inelastic scattering (DIS) of longitudinal polarized muons on longitudinal
polarized protons. ~EMC, ~using also ~the earlier ~data of SLAC
$^2$, came to
the surprising and contraintuitive conclusion that quarks are carrying
a small part of the proton spin projection in the polarized proton.
This statement results in excitement of one part of particle physics
community and to disappointment of the other. Many ideas, how to overcome
this problem were suggested, sometimes new and interesting, sometimes,
however, misleading. As a result of these investigations we understand
now much more not only about proton spin structure, but also about some
others connected with its issues, and this understanding will be with us
forever.

In my talk I will try to lay stress on this new understanding, which
arised in the last years, taking in mind the aphorism: \lq\lq The
result of calculation is not the number, but the understanding \rq\rq .
\vspace{5mm}


\noindent
{\large\bf 2. General Relations} $^{\ast}$
\vspace{3mm}

The polarized deep inelastic $e(\mu)$-nucleon scattering cross section
is determined by imaginary part of the forward virtual photon-nucleon
scattering amplitude $T^a_{\mu\nu} (p, q)$, antisymmetric in $\mu, \nu$.
The general form of $ImT^a_{\mu\nu}(p, q)$, following from gauge, $P$
and $T$ invariance is
\be
   Im T^a_{\mu\nu}(p, q) = \frac{2\pi}{m} \varepsilon_{\mu\nu\lambda\sigma}
   q_{\lambda} \left[ s_{\sigma} \left(G_1 + \frac{\nu}{m^2} G_2 \right)
   - (sq) \frac{p_{\sigma}}{m^2} G_2 \right] ~,
\ee
where $p$ and $q$ are the nucleon and virtual photon momenta, $s$ and $m$
-- are the nucleon spin and mass. The structure functions $G_1$ and $G_2$
depend on two invariants, $q^2 = -Q^2 < 0$ and $\nu = pq$. When $Q^2$ and
$\nu$ are large enough, but their ratio $x = Q^2/2\nu$ is fixed, $0 < x < 1$
(so called scaling limit), the scaling relations take place:
\be
\begin{array}{rr}
   (\nu / m^2) G_1 (\nu, q^2) ~ = ~ g_1 (x, Q^2)   \\ [2mm]
   (\nu / m^2)^2 G_2 (\nu, q^2) ~ = ~ g_2 (x, Q^2)
\end{array}
\ee

In the parton model $g_1$ and $g_2$ are functions of only scaling variable
$x$, in QCD a smooth (logarithmic) dependence of $g_1, g_2$ on $Q^2$
appears. In the parton model $g_1(x)$ can be represented through quark
distributions in the longitudinally --- along the beam polarized nucleon
\be
   g_1 (x) = \sum_{i=u,d,s,...} e^2_i [ q_{i+} (x) - q_{i-} (x) ] ~,
\ee
\underline{~~~~~~~~~~~~~~~~~~~~~~~~~~~~~~~~~~~~~}\\
\small
$^{\ast}$ For more details see $^3$.

\newpage
\normalsize
\noindent
where $q_{i+}(x)$ and $q_{i-}(x)$ are distribution of quarks in the
nucleon, polarized along or opposite to the nucleon spin, $e_i$ -- are
the quark charges. In the parton model $x$ has the meaning of the part
of the nucleon momentum, carried by quark in the infinite momentum frame
(or for fast moving nucleon, e.g. in the virtual photon-nucleon c.m.s.
at $Q^2 \to \infty$). Therefore, the integrals
\be
   \Delta q_i = \int\limits_0^1 d x [ q_{i+} (x) - q_{i-} (x) ]
\ee
have simple physical meanings: they are equal to the parts of nucleon spin
projection, in longitudinally polarized nucleon carried by the quarks of
flavour $i = u, d, s, ...$ .

The structure functions $G_1, G_2$ can be
expressed in terms of the virtual photon absorption cross section
$\sigma_{3/2}$ and $\sigma_{1/2}$ as well as the quantity $\sigma_I$
describing the transition from transverse to longitudinal virtual photon
polarization (on vice versa) in the forward scattering amplitude.
The cross sections $\sigma_{3/2}$ and $\sigma_{1/2}$ correspond to the
projections $3/2$ and $1/2$ of the total photon nucleon spin upon the
photon momentum direction.
\be
\begin{array}{rr}
    \sigma_{1/2} - \sigma_{3/2} = \frac{8\pi^2 \alpha}{\sqrt{\nu^2 - m^2q^2}}
    \frac{1}{m^2} (\nu G_1 + q^2 G_2)\\ [2mm]
  \sigma_I = \frac{4\pi^2 \alpha}{\sqrt{\nu^2 - m^2q^2}}
    \frac{Q}{m} (G_1 + \frac{\nu}{m^2} G_2)\\
\end{array}
\ee
In turn, $\sigma_{3/2}, ~~\sigma_{1/2}$ and $\sigma_I$ are proportional
to the $s$-channel helicity amplitudes $Im~T_{1, -1/2; 1, -1/2}, ~~
Im~T_{1, 1/2; 1, 1/2}$ and $Im~T_{1, 1/2; 0, -1/2}$, where the first two
indeces correspond to final photon and nucleon helicities, and the second
two to initial ones. The interference cross section $\sigma_I$ satisfies
the inequality
\be
    | \sigma_I | ~~ < ~~ \sqrt{\sigma_T \sigma_0} ~ =  ~ \sqrt{R} \sigma_0 ~ ,
\ee
where $\sigma_T$ and $\sigma_0$ are the absorption cross sections of
transverse and longitudinal virtual photons in nonpolarized scattering,
$\sigma_T = (\sigma_{3/2} + \sigma_{1/2})/2, R = \sigma_0/\sigma_T$.

The calculations of the structure functions in QCD are performed in the
framework of the operator product expansion (OPE). The operators are
classified according to twist ~=~ dimension ~-~ spin. The expansion in
twists is equivalent to expansion in inverse powers of $Q^2$ at fixed $x$.
This expansion starts with twist 2 terms in the case of $G_1$, and with
twist 3 terms in the case of $G_2(x, Q^2)$, corresponding of scaling
behavior of $g_1(x, Q^2), g_2(x, Q^2)$. In perturbative QCD only the
evolution with $Q^2$ of the structure functions can be determined, the
values of the structure functions at some fixed $Q^2 = Q^2_0$ are
taken from the experiment and used as an input for evolution equations.

The goal of all performed till now experiments was to measure the structure
function $g_1(x, Q^2)$. This goal was achieved in the following way.
The experimentally measurable quantity is the asymmetry
\be
   A ~ = ~ \frac{\sigma^{\downarrow \uparrow} - \sigma^{\uparrow \uparrow}}
                {\sigma^{\downarrow \uparrow} + \sigma^{\uparrow \uparrow}}
\ee
where $\sigma^{\downarrow \uparrow}, \sigma^{\uparrow \uparrow}$
correspond to the cases, when the spins of longitudinal polarized $\mu(e)$
and nucleon are antiparallel or parallel. The asymmetry $A$ is
related to
\be
     A_1 ~ = ~ (\sigma_{1/2} - \sigma_{3/2}) / (\sigma_{1/2} + \sigma_{3/2}) ~
,
     ~~~~~~~~ A_2 ~ = ~ \sigma_I / \sigma_T
\ee
by
\be
   A ~ = ~ D (A_1 + \eta  A_2)
\ee
where
\be
   D = \frac{y(2-y)}{y^2 + 2(1-y)(1+R)} ~ , ~~~~~~
   \eta = \frac{Q}{E} ~ \frac{2(1-y)}{y(2-y)}
\ee
$y = E - E', E$ and $E'$ are in initial and final $\mu(e)$ energies.
It can be shown using the inequality (6), that the second term in (9)
is small in practically interesting cases. Then the measurable quantity
is proportional to $A_1$ and the latter determines $g_1(x, Q^2)$
through the relation
\be
   g_1 (x, Q^2) = \frac{F_2 (x, Q^2) A_1 (x, Q^2)}{2x(1+R)} ~ ,
\ee
where $F_2(x, Q^2)$ is the structure function for nonpolarized DIS.
Before going to new developments in the theory of polarized structure
functions, I recall some facts which should be well known. However,
sometimes a misunderstanding occurs even here.
\vspace{5mm}


\noindent

{\bf 3. Small $x$ Domain -- Regge Behavior}
\vspace{3mm}

Consider the spin-dependent virtual photon-nucleon forward scattering
amplitude
\be
   T^a_{\mu\nu}(p,q) = \frac{2\pi}{\pi} \varepsilon_{\mu\nu\lambda\sigma}
   q_{\lambda} \left\{ s_{\sigma}S_1 (\nu, q^2) + \frac{1}{m^2}
   \left[ \nu s_{\sigma} - (sq) p_{\sigma} \right] S_2 (\nu, q^2)
   \right\} ~ ,
\ee
\be
   G_1 (\nu, q^2) ~ = ~ Im~S_{1,2} (\nu, q^2) ~ .
\ee
The functions $S_{1,2}(\nu, q^2)$ satisfy the crossing--symmetry relations
\be
   S_1 (\nu, q^2) ~ = ~ S_1 (-\nu, q^2) ~, ~~~~~~
   S_2 (\nu, q^2) ~ = ~ -S_2 (-\nu, q^2) ~ .
\ee
It can be shown, that at $\nu \to \infty$ and $Q^2 = Const$
(or $x \to 0, Q^2 = Const) ~ S_1 $  is contributed by Regge poles,
satisfying the condition
\be
    G (-1)^T \sigma ~ = ~ -1 ~ ,
\ee
where $G$ is $G$-parity, $T$ - is the isospin and $\sigma$ is the
signature. Then in accord with (12), (14) $\sigma = -1$ and the lowest
Regge trajectory contributing to $S_1$ are $a_1 (A_1)$ and $f_1 (D)$
meson trajectories
\be
   S_1 (\nu, q^2 ) \approx \beta_{a_1} (q^2) \nu^{\alpha_{a_1} - 1} ~ ,  ~~~~
   g_1(x) \sim x^{-\alpha_{a_1}}  ~ .
\ee
The intercept $\alpha_{a_1}$ is known not quite well, certainly it is
negative and about $\alpha_{a_1} \approx -(0.0-0.3)$. The same intercept is
expected for $f_1$ trajectory.

Up to the small contribution $\lsim \nu^{-1.5}$, arising from $S_1$
(see eq.(5)), the function $S_2$ is proportional to the helicity amplitude
$T_{1, 1/2; 0, -1/2}$. The Regge poles do not contribute to the latter.
This statement directly follows from the factorization theorem for Regge
poles. Indeed, the interaction vertex of virtual $\gamma$ with any Regge
pole in this case is a scalar, constructed from vectors $ {\bf q},
{\bf e}_L,  {\bf e}_T$,
 linear in the last two, what is impossible. For this reason
$S_2 (\nu, q^2)$ at large $\nu$ is determined by branch points
contributions (three pomeron cut etc.)
\be
   S_2 (\nu, q^2) \approx \beta_c (Q^2)/ln^5\nu + \sum_{i=P', A_2}
   \beta_i (Q^2) \nu^{\alpha_i - 1} /ln \nu + ...  ~~.
\ee
\vspace{3mm}


\noindent
{\bf 4. The Parton Model}
\vspace{3mm}

As was already mentioned, the spin dependent structure function $g_1(x)$
is represented in the parton model in terms of quark distributions by
eq.3 in comparison with the representation of nonpolarized structure
function
\be
   F_2 (x) = x \sum_i e^2_i [ q_{i+} (x) + q_{i-} (x) ] ~ .
\ee
The standard parametrization of quark distributions at some normalization
point $Q^2 = Q^2_0$ (usually $Q^2_0 \sim 5 ~GeV^2$) used in the descripton
on the nonpolazed DIS data is obtained by matching the behavior at small $x$
(Regge domain) with the behavior at large $x, 1 - x \ll 1$, following
from the quark counting rules:
\be
   q(x) \equiv q_+ (x) + q_- (x)  = Ax^{-\alpha} (1 - x)^{\beta} ~ .
\ee
It is expected that for valence quarks $\alpha \approx 0.5$ ($\rho$ -
intercept), $\beta \approx 3.0$, for sea quarks $\alpha \approx 1.0$
(experimentally $\alpha \approx 1.2$  is more preferable), $\beta
\approx 5$.

A more refined parametrization, which accounts for the fact that
$q_+(x)$ and $q_-(x)$ behave differently at $x \to 1$ ~
$q_-(x) / q_+(x) \sim (1-x)^2$ was suggested by Brodsky
(see the review $^4$).

The same parametrization (19) can be applied to spin dependent quark
distributions
\be
   q_+ (x) - q_-(x) ~ = ~ Bx^{-\alpha'} (1-x)^{\beta'} ~ ,
\ee
where $\alpha' = \alpha_{a_1} \approx -(0.0-0.3), \beta' \approx 3$
for valence and $\beta' \approx 5$ for sea quarks.

The parametrizations (19), (20) results in an interesting inequality in the
case of strange (sea) quarks $^{5, 6}$. For strange quarks (19), (20)
reduce to
\be
 \begin{array}{cc}
    x s (x)  \equiv x [ s_+ (x) + s_- (x) ] \approx A_s (1 - x)^{\beta_s} ~ ,\\
    s_+ (x) - s_- (x) \approx B_s (1 - x)^{\beta_s} ~ ,
 \end{array}
\ee
where it was accepted for simplicity that $\alpha = 1.0, \alpha' = 0.$
Since $s_{\pm} (x) \ge 0,$ we have an inequality
\be
    A_s ~~ \ge ~~ | B_s | ~ ,
\ee
which results in
\be
   V_{2,s} \equiv \int\limits^1_0 x s (x) dx \ge \int\limits^1_0
   dx[s_+(x) - s_-(x)] \equiv \Delta s
\ee
in the notation (4). The inequality (23) states, that, if the parametrization
(23) is correct at some $Q^2_0$, then at this $Q^2_0$ the part of proton spin
projection carried by strange quarks cannot exceed the part of proton
momentum carried by them. It is clear that small deviation from $\alpha =
1.0, \alpha' = 0.0,$ say $\Delta \alpha_{T} = -0.2, \Delta \alpha' = 0.3$
cannot seriously violate the inequality (23).
\vspace{5mm}


\noindent
{\bf 5. Sum Rules in the Parton Model}
\vspace{3mm}

The sum rules for spin dependent structure functions of deep inelastic
scattering follows from space-time representation of amplitudes in
terms of current commutators:
\be
   Im~T_{\mu\nu}^a (x) = \frac{1}{4} < p, s | [ j_{\mu} (x),
   j_{\nu} (0) ] | p, s >_{antisym.}  ~.
\ee
Current commutators satisfy the causality condition
\be
   [ j_{\mu} (x), j_{\nu} (0) ] ~ = ~ 0 ~~~~~~\mbox{at}~~~~ x^2 < 0  ~.
\ee
It can be shown that the domain near light cone corresponds to large
$Q^2 ~^7.$ The sum rules can be derived by considering eq.(24) in the
limit $x^2 \to 0, x_0 \to 0$ and using equal time commutation relations
for calculation of current commutators, expressed in terms of free quark
fields. Accounting for $u, d, s$ quarks we have from consideration of
space-like components $\mu = i, \nu = k$ in (24):
\be
   Im ~T_{ik} (x)_{x_0 \to 0} = - \varepsilon_{ikl} < p, s
   \mid \frac{1}{3} \left[ j^3_{5l} (0) + \sqrt{\frac{1}{3}} j^8_{5l}
   (0) \right] + \frac{2}{9} j^0_{5l} \mid p, s > ~ ,
\ee
where $j^3_{5l}, j^8_{5l}$ and $j^0_{5l}$ are isovector, octet  and
singlet (in SU(3) flavour symmetry) axial currents. The matrix elements
in the r.h.s. of (26) are proportional to the nucleon coupling constants
with axial currents. The isospin symmetry determines the proton matrix
element of isovector current
\be
   < p, s \mid j^3_{5l} (0) \mid p, s >_p ~ =  -2mg_As_l ~ ,
\ee
where $g_a = 1.25$ is the neutron $\beta$-decay axial coupling constant.
On the other side, using the Fourier transformation of eq.1, the l.h.s.
of (26) can be represented in terms of the structure functions
(for details see $^3$). In this way the famous Bjorken sum rule $^8$
arises
\be
   \int\limits^1_0 [g^p_1 (x) - g^n_1 (x) ] dx = \frac{1}{6} g_A ~.
\ee
The Ellis-Jaffe sum rule $^9$ for octet current follows if SU(3)
flavour symmetry for baryonic octet $\beta$-decays is supposed
\be
   \int^1_0 g^8_1 (x) dx ~ = ~ \frac{1}{24} (3F-D) ~ ,
\ee
where $F$ and $D$ are $\beta$-decay axial coupling constants in the
baryonic octet.

In a similar way, by considering the behaviour of $ImT^a_{\mu\nu} (x)$ (24)
near the tip of light cone the Burkhardt-Cottingham $^{10}$
sum rule for the structure function $g_2(x)$ can be derived
\be
    \int^1_0 g_2 (x) dx ~ = ~ 0  ~.
\ee
In the derivation (30) no current algebra is used, only the causality
condition (25) and the hypothesis of absence of nontractable
singularities in the amplitude at the tip of the light cone are
imposed.

The sum rule (30) looks like a superconvergent sum rule, as it was
originally derived in Ref.$^{10}$. This is not the case, however.
Indeed, in accord with (17) we can write the subtractionless dispersion
relation for $S_2 (\nu, q^2 )$
\be
   S_2 (\nu, q^2) ~ = ~ 4\nu \int\limits^{\infty}_{Q^2/2}
   \frac{G_2 (\nu', q^2)}{\nu'~^2 - \nu^2} d \nu'  ~.
\ee
If $S_2 (\nu, q^2)$ would decrease faster than $1/\nu$ at $\nu \to \infty$,
then from (31) the sum rule would follow
\be
    \int\limits^{\infty}_{Q^2/2} d \nu' G_2 (\nu', q^2) ~ = ~ 0 ~ ,
\ee
which  goes into (30) in the limit $Q^2 \to \infty$. But due to
contributions of pomeron, $P'$ and $A_2$ cuts, $S_2 (\nu, q^2)$
decreases slower than $1/\nu$ (see (17)) and (32) is invalid. That
means, that there are higher twist contributions to $g_2 (x, Q^2)$
strongly increasing at small $x$ in comparison with lowest twist one,
of the order $(1/Q^2)(1/x^2)ln^{-n}x, ~ n \ge 5.$ (The sum rule (30)
validity was confirmed by QCD sum rule calculation in the lowest order
of the operator product expansion in $p^2$ - nucleon virtuality $^{11}$.)

Finally, I discuss the Gerasimov, Drell, Hearn (GDH) sum rule $^{12}$.
GDH sum rule refers to the scattering of the real photon on polarized
nucleon. But, as we will see, it is closely related to the sum rules for DIS.

In the case of real photon only one amplitude $S_1 (\nu, 0)$ survives in (12),
because the kinematical factor in front of the
other - $S_2(\nu, 0)$  is
zero. The function $S_1(\nu, 0)$ satisfies the unsubtracted dispersion
relation
\be
    S_1 (\nu, 0 ) ~ = ~ 4 \int\limits^{\infty}_0 \nu' d \nu'
    \frac{G_1 (\nu', 0)}{\nu'~^2 - \nu^2}  ~ .
\ee
Let us go to the limit $\nu \to 0$ in this relation. According to the
F.Low theorem, the terms, proportional to zero and first power in the
photon frequency in the photon-nucleon scattering amplitude are
expressed via the static characteristics of the nucleon. The direct
calculation of Feynman graphs gives
\be
   S_1 (0, 0)_{p, n} ~ = ~ - \kappa ^2_{p,n} ~ ,
\ee
where  $\kappa_{p,n}$ are the anomalous proton and neutron magnetic moments.
{}From (33), (34) the GDH sum rule follows
\be
   \int\limits^{\infty}_0 \frac{d\nu}{\nu} G_{1, p, n} (\nu, 0)
    ~ = ~ - \frac{1}{4} \kappa ^2_{p, n} ~ .
\ee
The essential point is that at $Q^2 = 0$ and $\nu \to 0$ the spin dependent
forward Compton amplitude is a constant and has no nucleon pole. That
means that in the l.h.s. of the sum rule (35) the elastic contribution is
absent --- only inelastic processes contribute. That is why the GDH sum
rule is very nontrivial.
\vspace{5mm}


\noindent
{\bf 6. QCD Corrections to Sum Rules}
\vspace{3mm}

In what follows we use the notation
\be
    \Gamma_{p, n} (Q^2) ~ = ~ \int\limits^1_0 g_{1,p,n} (x, Q^2) d x ~ .
\ee
In the parton model
\be
    \Gamma_{p, n}  ~ = ~ \frac{1}{12} \left( \pm a_3 + \frac{1}{3} a_8
    + \frac{4}{3} \Sigma \right) ~ ,
\ee
where, for 3 flavours in notation of (4)
\be
  \begin{array}{cc}
   a_3 ~ = ~ \Delta u - \Delta d ~; ~~~~~~ a_8 ~ = ~ \Delta u + \Delta d
   - 2 \Delta s ~ ,\\
   \Sigma ~ = ~ \Delta u + \Delta d + \Delta s
  \end{array}
\ee
and $\Sigma$ has the meaning of part of the nucleon spin projection
carried by quarks. The sum rules, presented in Sec.4 in the supposition
of SU(3) flavour symmetry in baryonic $\beta$-decays result in
equations
\be
    a_3 ~ = ~ g_A ~ ,    ~~~~~~~~ a_8 ~ = ~ 3F - D ~ .
\ee
In QCD eq.37 is modified by account of $\alpha_s$-corrections. The first
order corrections in $\alpha_s$ were calculated in $^{13-17}$, the
$\alpha^2_s$ and $\alpha^3_s$  corrections to the octet part of the sum
rule in $^{18}$, the $\alpha^2_s$ corrections to the singlet part in
$^{19}$. With QCD corrections, calculated in $\overline{MS}$ regularization
scheme the sum rules takes the form (for 3 flavours)
$$   \Gamma_{p,n} (Q^2) = \frac{1}{12} \left\{ \left[ 1 -
   \frac{\alpha_s (Q^2)}{\pi} - 3.6 \left( \frac{\alpha_s (Q^2)}{\pi}
   \right)^2 - 20 \left( \frac{\alpha_s (Q^2)}{\pi}\right)^3 \right]
   \times \right.   $$
$$   \left. \times \left( \pm a_3 + \frac{1}{3} a_8 \right)
    + \frac{4}{3} \left[ 1 - \frac{1}{3} \frac{\alpha_s (Q^2)}{\pi}
    - 1.1 \left( \frac{\alpha_s (Q^2)}{\pi} \right)^2 \right]
    \Sigma \right\} - $$
\be
    - \frac{N_f}{18\pi} \alpha_s (Q^2) \Delta g (Q^2)
\ee
where $N_f = 3$ is the number of flavours,
\be
   \Delta g(Q^2) = \int^1_0 dx [g_+(x) - g_-(x)] ~,
\ee
$g_{\pm}(x)$ are the distrubutions of gluons in the longitudinally
polarized nucleon with spin projections along or opposite to the nucleon spin.
$\Delta g$ has the meaning of the part of nucleon spin projection
carried by gluons. In eq.(40) $a_3, a_8$ and $\Sigma$ are $Q^2$
independent.

A physically new point in eq.40 is the appearance of gluonic
contribution to the sum rule. Although the last term in (40) is an
$\alpha_s$ correction and naively it would be expected to vanish in the
limit $Q^2 \to \infty$, in fact it does not, because the $\Delta g(Q^2)$
anomalous dimension is equal to $-1$, i.e.
\be
   \Delta g (Q^2)_{Q^2 \to \infty} \sim ln Q^2
\ee
and
\be
   \lim_{Q^2 \to \infty} \alpha_s (Q^2) \Delta g (Q^2) = Const ~.
\ee
 For this reason the last term in (40) cannot be considered as a correction
vanishing at $Q^2 \to \infty$, like any other $\alpha_s (Q^2)$
corrections (e.g. the $\sim \alpha_s$ terms in the figure bracket in (40),
where $a_3, a_8$ and $\Sigma$ have zero anomalous dimensions).

On the other side from the nucleon spin projection conservation it follows
that
\be
   \frac{1}{2} \Sigma + \Delta g (Q^2) + L_z (Q^2)  = \frac{1}{2} ~,
\ee
where $L_z$  is the orbital momenta contribution. Therefore,
the logarithmically increasing with $Q^2 ~ \Delta g (Q^2)$ means that
$L_z (Q^2) \sim ln Q^2 $  at large $Q^2$, is negative and compensates
$ \Delta g (Q^2)$. As a direct consequence the quark model of nucleon with
quarks in $S-$states cannot work at high $Q^2$.

There was a wide discussion in the past years if gluonic contribution
$\Delta_g \Gamma_{p,n}$ to $\Gamma_{p,n}$ (the last term in eq.(40) )
is uniquelly defined theoretically or is not $^{17, 20-32}$. The problem
is that gluonic contribution to the structure functions, described by
imaginary part of the forward $\gamma_{virt}$ - gluon scattering amplitude
(Fig.1) is infrared dependent. Since in the infrared domain the gluonic
and sea quark distributions \\
\vspace{12.5mm}

\noindent
\begin{tabular}{p{90mm}p{55mm}}
 & \underline{\bf Fig. 1}\\
 & The photon-gluon scattering diagrams, the wavy, dashed and solid lines
correspond to virtual photons, gluons and quarks.
\end{tabular}
\vspace{12.5mm}

\noindent
are mixed and their separation depends on the infrared
regularization scheme, a suspicion arises that $\Delta_g \Gamma_{p, n}$
can have any value. This suspicion is supported by the fact that in the lowest
order in $\alpha_s$ the terms, proportional to $ln Q^2$ are absent in
$\Delta_g \Gamma_{p,n}$ and this contribution looks like next to leading
terms in nonpolarized structure functions, where such an uncertainty
is well known.

In order to discuss the problem consider the gluonic contribution
$g_{1p} (x, Q^2)_{gl}$ to the proton structure function $g_{1p} (x, Q^2)$,
described by evolution equation
\be
   g_{1p} (x, Q^2)_{gl} = N_f \frac{<e^2>}{2}  \int^1_x
   \frac{dy}{y} A \left(  \frac{x}{y} \right) [g_+ (y) - g_- (y)] ~,
\ee
where the asymmetry $A(x_1)$ is determined by the diagrams of Fig.1.
The calculation of the asymmetry $A(x_1), x_1 = -q^2/2pq$ results in the
appearance of integrals
$$
  \int \frac{d^2 k_{\perp}}{\left(  k^2_{\perp} - x_1 (1 - x_1)
   p^2 + m^2_q  \right)^n} ~ ,   ~~~~~~ n = 1, 2 ~ ,
$$
which are infrared dependent. To overcome this problem it is necessary
to introduce the infrared cut-off (or infrared regularization), to
separate the domain of large $k^2_{\perp}$, where perturbative QCD
is reliable, from the domain of small $k^2_{\perp}$. The contribution
of the latter must be addressed to noncalculable in perturbative QCD
parton distribution. Such a procedure is legitimate because of
factorization theorem, which states that the virtual photoabsorption
cross section on the hadronic target $ h, \sigma^{\gamma}_h (x, Q^2)$
can be written down in the convolution form
\be
   \sigma^{\gamma}_h (x, Q^2)  = \sum\limits_i \sigma^{\gamma}_i
   (x, Q^2, M^2) \otimes f_{i/h} (x, M^2) ~ ,
\ee
where $\sigma^{\gamma}_i$ is the photoproduction cross section on
$i^{th}$ parton ($ i = q, \bar q, g$), ~$f_{i/h}$ are the parton
distributions in a hadron $h, \otimes$ stands for convolution. Both
$\sigma^{\gamma}_i$ and $f_{i/h}$ depend on the infrared cut-off $M^2$,
but the physical cross section $\sigma^{\gamma}_h$ is cut-off
independent. The variation of $M^2$ corresponds to redistribution
among partons: the trade of gluon for sea quarks.

As follows from (45)
\be
   \Delta_g \Gamma_p = N_f \frac{<e^2>}{2}  \Delta  g
    \overline{A(M^2)} ~ ,
\ee
where
\be
   \overline{A(M^2)} = \int^1_0 dx_1 A (x_1, M^2) ~.
\ee
The convenient way is to introduce cut-off in $k^2_{\perp}$
\be
   k^2_{\perp}  ~ > ~ M^2 (x_1, p^2)
\ee
Generally, $M^2$ may depend on  $x_1$ and $p^2$. For example,
the cut-off in quark virtuality in Fig.1 $-k^2 > M^2_0$
corresponds to the form of (49) with $M^2 = (1-x_1)(M^2_0 + p^2 x_1)$
if $x_1 < - M^2_0/p^2$.

In the calculation of the diagrams of Fig.1 it is reasonable to neglect
the light quark masses in comparison with gluon virtuality $p^2$, since we
expect that $\mid p^2 \mid$ is of order of characteristic hadronic masses,
$\mid p^2 \mid \sim 1 ~GeV^2~^{20}$. Then introducing the infrared cut-off (49)
we have $^{31}$
\be
   \overline{A(M^2)} = - \frac{\alpha_s}{2\pi} \left\{ 1 - \int\limits^1_0
   d x_1 (1 - 2x_1) [ln r - r] \right\} ~ ,
\ee
where
\be
   r = \frac{x_1 (1 - x_1) p^2}{x_1 (1-x_1) p - M^2 (x_1, p^2)} ~.
\ee
If $M^2 = Const $ - a rectangular cut-off in $k^2_{\perp}$ - , the integral
in (50) vanishes (the integrand is antisymmetric under substitution
$x_1 \to 1 - x_1$). Then $\bar A = -\alpha_s/2\pi$ and we obtain the
gluonic contribution to $\Gamma_{p, n}$ (40). However, other forms of
$M^2(x_1, p^2)$ result in different values of $\overline{A(M^2)}$,
what support the
claim $^{21}$ that $\overline{A(M^2)}$ is cut-off dependent.
Even more, if we put
$p^2 = 0, m^2_q \ne 0$ - the standard regularization scheme in the
calculation of nonpolarized deep inelastic scattering - we will find
$\bar A = 0 ~^{29}$. This result is, however, nonphysical because
the compensation of $-\alpha_s/2\pi$ term in $\bar A$ arises from soft
non-perturbative domain of $k^2_{\perp} \sim m^2_q$, which must be attributed
to sea quark distribution.

Although generally $\bar A $, as well as $\Delta g$ and $\Delta_g \Gamma_p$,
are infrared cut-off dependent, a special class of preferable cut-off's can be
chosen. In OPE the mean asymmetry $\bar A$ is proportional to the one-gluon
matrix element of axial current
\be
   \Gamma_{\mu \lambda \sigma}(0, p, p) ~ = ~ < g_1 \varepsilon_{\lambda}
   \mid  j_{\mu 5} (0) \mid  g_1 \varepsilon_{\sigma} >  ~,
\ee
at zero momentum transfer. It can be shown $^{31}$, that this quantity is
proportional to the divergency $l_{\mu } \Gamma_{\mu \lambda \sigma}
(l, p_1, p_2), l = p_2 - p_1$, in the limit $l^2 \to 0$, i.e. to the one gluon
matrix element of axial anomaly. The standard expression for axial anomaly
corresponds to $M^2 = Const$ and $\bar A = - \alpha_s/2\pi$ (although it was
demonstrated $^{31}$, that the same infrared dependence persists here also).
Therefore, the cut-off in $k^2_{\perp} < M^2 = Const$ is preferable, since it
preserves the standard form of axial anomaly. In this case the axial
anomaly can be considered as a local probe of gluon helicity. Some care,
however, is necessary, when the calculations with this cut-off are compared
with HO calculation in nonpolarized scattering, where, as a rule, another
regularization procedure $-p^2 = 0, m^2_q \ne 0$ - is used.

The gluonic contribution to the spin dependent proton structure function
$g_{1p} (x, Q^2)_{gl} $ can be found experimentally by measuring inclusive
two jets production in polarized DIS $^{20}$. However, only large -
$k^2_{\perp}$ component of Fig.1 diagrams can be determined in this way, the
small - $k^2_{\perp}$ component is unmeasurable, since in this case it is
impossible to separate one jet from two jets events $^{31}$. This fact is in
complete accord with the formulated above statement about the arbitrariness
of infrared cut-off.

What numerical value of gluonic contribution to $\Gamma_{p, n} (Q^2)$
can be expected? Naively, we can say, that since gluons are carrying about
50\% of proton momentum $< x >_{gl} \simeq 0.5$ $^{\ast}$ ,
 they can carry the same amount of proton spin projection, $\Delta g
\approx 1/4 - 1/2$ (see (44) ). The same conclusion follows from the
simplest parametrization of gluon distribution like (19), (20), or from the
 parametrization, suggested by Brodsky $^4$. If $\Delta g \approx 1/2$
at $ Q^2 = 10 ~GeV^2$ ($\alpha_s (10 ~GeV^2) \approx 0.25$ at $\Lambda_{QCD}
= 200 MeV$) then
\be
   \Delta_g \Gamma_{p, n} \approx 0.0066 ~ .
\ee
Earlier Brodsky and Schmidt $^{33}$ basing on the model of bound-state
nucleon and counting rules suggested a different parametrization and
obtained $\Delta g \approx 1.2$ at $Q^2 \sim 1 ~GeV^2.$  Up to terms of
order $\alpha^2_s$ the perturbative evolution of $\Delta g (Q^2)$ with
$Q^2$ is given by $^{23, 35}$
\be
    \begin{array}{cc}
   \Delta g (Q^2) = \frac{\alpha_s (\mu^2)}{\alpha_s (Q^2)}
   \left\{ 1 + \frac{2N_f}{b\pi} \left[ \alpha_s (Q^2) - \alpha_s
   (\mu^2) \right] \right\} \Delta g (\mu^2) + \\
+ \frac{4}{b}  \left[   \frac{\alpha_s (\mu^2)}
{\alpha_s (Q^2)} - 1  \right] \Sigma (\mu^2) ~ ,
    \end{array}
\ee
where $b = 11 - (2/3) N_f = 9, \mu^2 $ - is the normalization point.
As follows from (54), if the value $\Delta g = 0.25$ is accepted at
$\mu^2 = 1 GeV^2$, then $\Delta g (10 GeV^2) \approx 0.6$ (at
$\Lambda_{QCD} = 200 MeV $ and $\Sigma (\mu^2) = 0.59$ - see Sec.6).
In order to have higher values of $\Delta g ~(10 GeV^2)$ it is necessary
to start from lower normalization point. \\
\underline{~~~~~~~~~~~~~~~~~~~~~~~~~~~~~~~~~~~~~} \\
\small
\begin{tabular}{p{6mm}p{140mm}}
$^{\ast}$
& This number refers to typical $Q^2$ in DIS experiments, $Q^2 \sim
5 ~GeV^2$.  At $Q^2 \gsim 2 ~GeV^2 < x >_{gl}$ only slighly depends on
$Q^2$, but is decreasing at low $Q^2$ up to $< x >_{gl} \approx 0.3 $ at
$Q^2 = 0.3 ~GeV^2 ~ ^{34}$. Since $\Delta g (Q^2)$ increase as $ln~Q^2$ at
high $Q^2$, the conclusion $\Delta g \lsim 0.5$ is valid if we assume, that
gluonic distribution is described by equations like (19), (20) at $Q^2 \sim
 2 - 10 ~GeV^2.$
\end{tabular}

\newpage
\normalsize
\noindent
E.g. $\Delta g ~(10 GeV^2) = 3$
requires $\Delta g = 1$  at $\mu^2 = 0.3 GeV^2 ~^{35}$. The large value of
$\Delta g$ in the latter case is achieved due to increasing of
 $g_+(x) - g_-(x)$ at small $x$ and the problem arises if such increasing is
compatible with experiment and does it proceeds in the domain of $x,$
measured in the existing experiments. There are different opinions about this
subject $^{30, 31, 35-37}$. Perhaps the problem must be reinvestigated in
the light of new experimental data.

It is interesting to mention that in the ratio of Gross-Llewellyn-Smith and
Bjorken sum rules the perturbative QCD corrections cancel up to $\alpha^2_s
{}~ ^{18}$ (see also the discussion of this point in $^4$). Therefore the
experimental examination of this prediction at moderate $Q^2$ could be also
a good check of perturbative QCD and the role of higher twist terms.

Let us turn now to twist four corrections to the sum rules (40). The general
theory of these corrections was developed by Shuryak and Vainstein $^{38}$.
According to this theory  the twist 4 corrections to (40) are proportional
to the one-nucleon matrix elements of the operators
\be
   < N \mid U_{\mu}^{S, NS} \mid N > = < N \mid g \bar q \tilde G^a_{\mu\nu}
   \gamma_{\nu} \frac{1}{2} \lambda^a (1, \tau_3) q \mid N > ~ ,
\ee
where $\tilde G^a_{\mu\nu}$ is the dual gluonic field tensor, $S(NS)$
are singlet (nonsinglet) in flavour operators, $1(\tau_3)$  in the r.h.s.
of (55) correspond to $S(NS)$. In (55) $q$ means $u$ and $d$ and the
contribution of strange quarks is omitted. An attempt to determine the
 matrix elements (55) in the framework of QCD sum rule 3-point function
calculations ~$^{39, 40}$ was done by Balitsky, Braun, Kolesnichenko
{}~$^{41}$.  Their final results are
\be
  \Delta_{twist 4} \Gamma_{p,n} =
   - \frac{1}{18} (4, 1) \frac{0.09 \pm 0.06}{Q^2} GeV^2 + \frac{2}{9}
   \frac{m^2}{Q^2 } \int\limits^1_0 x^2 g_{1, p, n} (x) dx ~.
\ee
The magnitude of twist 4 correction following from (56) is very small.  Even
with account of (negative) error it comprises $\sim -1 \cdot 10^{-3}$ at
$Q^2 = 10 GeV^2$ for proton as well as for neutron.

Unfortunately, I cannot believe that the result (56) is reliable for the
following reasons. In finding the vacuum expectation values induced by
external axial field - a quantity, which essentially determines the final
answer - the authors of ref.~$^{41}$  have really taken the octet field
instead of the singlet one and used the dominance of massless goldstones
($ \pi $ or $\eta$) which is incorrect for the singlet field case. For
the nonsinglet case the situation is, in principle better.
But, unfortunately, in
this case the main contribution, to the sum rule used by the authors,
comes from the highest order accounted term in the OPE - the term
$\sim m^2_0 < \bar{\psi} \psi >^2$ of dimension 8. It is possible that in this
case the contribution of unaccounted terms of OPE is essential or even
that the OPE series is divergent at characteristic values of the Borel
parameter $M^2 = 1 ~GeV^2$ used in the sum rule. On the physical side of the
sum rule the contribution of continuum comprises 80\% and the nucleon pole term
from which $\Gamma_{p-n}$ was obtained gives only 20\%. This also spoils the
accuracy of the calculations. Finally, the result of ~$^{41}$ (unlike the other
results obtained by the QCD sum rule) depends on the ultraviolet cut-off.
This circumstance introduces a noncontrollable uncertainty into the
calculation.
\vspace{5mm}

\noindent
{\large\bf 7. Nucleon Axial Coupling Constants}
\vspace{3mm}

As was demonstrated in the previous Sections, according to the sum rules
the integrals
$$
  \Gamma_{p,n} = \int g_{1,p,n}(x) d x
$$
are expressed through the nucleon matrix elements of isovector, octet and
singlet axial currents, $a_3, a_8$ and $\Sigma$. The magnitude of $a_3$,
determining Bjorken sum rule, is well known. According to isospin
invariance it is equal to the axial coupling constant in neutron $\beta$--
decay, $a_3 = g_A = 1.257 \pm 0.003 ~^{42}$. The theoretical accuracy
in this number is given by the accuracy of isospin symmetry, i.e. it is
of order 1\%. In a similar way $a_8$ is proportional to the matrix element
of octet axial current
\be
    a_8 = \sqrt{3} s_{\mu} < p, s \mid j_{\mu 5}^{(8)} \mid p, s > /2 m ~ .
\ee
This matrix element can be found if $SU(3)$ flavour symmetry in baryonic octet
$\beta$-decays is assumed. Then
\be
   a_8 = 3F - D = 0.59 \pm 0.02
\ee
and the numerical value in (58) corresponds to the best fit $^{43}$ of
neutron and hyperon $\beta$--decays in $SU(3)$ symmetry. The use of SU(3)
symmetry in determination of $a_8$ was questioned by Lipkin $^{44}$.
Indeed, the nondiagonal matrix elements enter neutron and hyperon $\beta$
--decays, while the diagonal one appears in (57). If $SU(3)$ symmetry is badly
violated, it is possible that the latter has nothing to do with the
formers. The accuracy of $SU(3)$ symmetry in the matrix elements of octet
axial current can be estimated in the following way $^{6}$. In $SU(3)$
symmetry the interesting for us combination $3F - D$  can be found from
any pair of $\beta$--decays. If $SU(3)$ is strougly violated we would
expect that the values, determined from various pairs of $\beta$--decays
are different. Experiments tell us $^{42}$
\be
   \begin{array}{rrrr}
   (3F - D)_{np, \Lambda p} = - 3 g_{A, np} + 6 g_{A, \Lambda p}
       = 0.537 \pm 0.09 ~,    \\
   (3F - D)_{np, \Sigma n} =  g_{A, np} + 2 g_{A, \Sigma n}
       = 0.577 \pm 0.034 ~,    \\
   (3F - D)_{\Lambda p, \Sigma n} =  (3/2) (g_{A, np} +  g_{A, \Sigma n})
       = 0.567 \pm 0.034 ~,    \\
    (3F - D)_{\Xi \Lambda} = 3 g_{A, \Xi \Lambda} = 0.75 \pm 0.15  ~.
   \end{array}
\ee
We see that all values (59) argee very well with (58) and, at least,
the value of $3F - D$ less than 0.5 is completely excluded.

The matrix element $a_8$ (57) was also determined theoretically by QCD
sum rule method $^{45}$. In this calculation no $SU(3)$ flavour symmetry
of baryonic $\beta$--decays was assumed. The result
\be
   a_8 ~ = ~ 0.5  ~\pm~ 0.2
\ee
is in agreement with (58).

To estimate the proton matrix element of singlet axial current (or $\Sigma$)
is a more complicated, till now nonsolved problem. Naively, by analogy with
a part of the proton momentum, carried by quarks, which is about 50\% at
$Q^2 \sim 5 - 10 ~GeV^2$, we expect, that $\Sigma \approx 0.5$. The same
estimate follows also from the assumption that strange sea is nonpolarized
(the original Ellis-Jaffe $^{9}$ assumption). Then $\Sigma \approx a_8 $
given by (58).

 On the other side, assuming that the Skyrme model is suitable for the
description of proton constituents, Brodsky, Ellis and Karliner $^{46}$
argue that $\Sigma$ is small of order $1/N_c$  in the limit of large
number of colours $N_c$. Indeed, in the Skyrme model the nucleon is built
from octet of goldstones, which is decoupled from singlet axial current
at $N_c \to \infty$. The Skyrme model, probably, gives a good description
of nucleon periphery. However, the description with this model of deep inside
nucleon structure, which is measured in deep inelastic scattering, is
questionable (see also $^{47}$).

In ref.48 an attempt was performed to calculate the nucleon matrix element
of singlet axial current $a_0 = \Sigma$ by QCD sum rules in the similar
way as it was done in the calculation of $a_8$ in $^{45}$. The main
difference with $a_8$ calculation is that in this case the anomaly plays
an important role in the vacuum expectation value induced by external
axial field, which essentially determines the final result. This problem
was overcome in $^{48}$, but nevertheless the result was negative:
an attemt to determine $\Sigma$ fails, since it was found that the OPE
breaks down for singlet axial current in longitudinal and transverse parts of
polarization operators and/or in the NNA vertex function. The physical
consequence of this consideration is that one should expect o noticeable
Okubo-Zweig-Iizuka (OZI) rule violation in the nonet of axial $1^{++}$
mesons.

The authors of Refs.24,49
tried to use the so called \lq\lq $U(1)$ Goldberger--Treiman\rq\rq  ~relation
\be
   F_{\eta_0} (0) g_{\eta_0 N N} = 2 m a_0
\ee
to determine $a_0$.
Unfortunately, no definite conclusion could be obtained in this approach
without serious additinal hypothesis. The reason for this comes from the
fact that due to anomaly $\eta_0$ in (61) is not a goldstone and not a
physical $\eta'$ meson. Perhaps, the troubles appearing in this approach
are of the same origin as in $^{48}$.
\vspace{5mm}
\newpage

\noindent
{\large\bf 8. The Experimental Data on $g_1(x)$ and Their Interpretation}
\vspace{3mm}

The \lq\lq Sturm und Drang\rq\rq ~period in the investigation of nucleon spin
structure started after appearance of famous EMC results $^{1}$, where
the structure function $g_{1p} (x)$ was measured in the interval $0.015
< x < 0.47$ at the mean $\bar Q^2 = 10.7 ~GeV^2.$ Combining their data with
earlier SLAC $^2$ data, where the measurements were
made at $0.1 < x < 0.65$
at $\bar Q^2 \approx 3 ~GeV^2$ and performing the extrapolation in the
domain $x < 0.015$ and $x > 0.65$, EMC obtained
\be
  \Gamma_p = 0.126 \pm 0.010 (statist) \pm 0.015 (system.) ~,
  ~~~\bar Q^2 = 10.7 ~GeV^2
\ee

Compare (62) with the simplest theoretical version --- the Ellis--Jaffe
sum rule, --- eq.40, where the gluonic contribution is neglected and it is
supposed that strange sea is nonpolarized, $\Delta s = 0$. Then $\Sigma
= a_8$ and using (58), we have at $\alpha_s (Q^2 = 10.7 ~GeV^2) = 0.25$
$(\Lambda_{QCD} = 200 ~MeV)$
\be
   \Gamma^{EJ}_p ~ = ~ 0.171 ~\pm~ 0.004 ~ ,
\ee
where the error comes from the error in (58) and from the uncertainty in the
$\alpha_s$ correction, which could be of the order of $\alpha^3_s$ term
(besides of uncertainty in $\Lambda_{QCD}$). The Ellis--Jaffe prediction
disagrees with (62) by 2.6 standard deviations.

Reject now the hypothesis $\Delta s = 0$, but still neglect gluonic
contribution.  Using experimental value (62) we can determine from (39),
(40), (58) the values $\Delta u, \Delta d, \Delta s$ and $\Sigma$.
\be
   \begin {array}{cc}
\Delta u = 0.785 \pm 0.06 ~, ~~~~ \Delta d = -0.47 \pm 0.06 ~,
{}~~~~ \Delta s = -0.14 \pm 0.06 ~ , \\
\Sigma = 0.17 \pm 0.17 ~ .
   \end{array}
\ee
The results (64) as originally was stressed in $^{1, 46}$ are very strange
and unexpected: the part of proton spin projection carried by quarks
is small and compatible
with zero, strange quarks are carrying a large amount of proton spin in
contradiction with the quark model and inequality (23). Indeed the
experimental data on DIS give $^{50}$
\be
   V_{2, s} ~ = ~ 0.026 ~ \pm ~ 0.006
\ee
and (23) is strongly violated.

The account of gluons does not improve the situation essentially, if
$\Delta g \approx 0.5$  at $Q^2 = 10 ~GeV^2.$ Using (53), we find that in this
case $\Delta s$ and $\Sigma$ increase
in comparison with (64) correspondingly by 0.02
and 0.07. If, we assume that $\Delta g (10 ~GeV^2) \approx 2$, then
\be
   \Delta s = - 0.06 \pm 0.06 ~; ~~~~~~ \Sigma = 0.42 \pm 0.17 ~,
\ee
the inequality (23) is fullfilled within the
errors and $\Sigma$ does not contradict
the naive expectations. The value $\Delta g (10 ~GeV^2) = 2 $ can be
achieved, if $\Delta g (1 ~GeV^2) = 1.2$ what was suggested in $^{33}$.
This solution means, however, that the constituent quark model would be
strongly violated at $Q^2 \sim 1 ~GeV^2$, since, according to (44) we
would have $L_z (1 ~GeV^2) \approx - 1.$

Leaving aside the discussion of the scenario with
large $\Delta g$, let us consider if it is possible to have large
$\mid \Delta s \mid$, violating the inequality (23). Qualitatively, the
problem can be formulated as follows: could it be, that in one experiment
we see a small number of strange quark pairs in the proton and a much
larger one in the other? The answer is $^{51}$: since the number of
$\bar s s$ pairs is not conserved, the question how many strange quark
pairs are in the proton is not correctly formulated until one specify
the operator, with which the proton strange content is measured.
The measurements of a part of proton momentum, carried by $s$--quarks
corresponds to the matrix element of the $s$--quark energy momentum
tensor operator, $V_{2s} \sim < p \mid \theta^s_{\mu \nu} \mid p > $,
the measurement of proton spin projection, carried by $s$--quark
corresponds to the $s$--quark axial current operator $\Delta s \sim
< p \mid j^s_{\mu 5} \mid p >$. The small value of $V_{2s}$
and a large of
$\Delta s$ would mean, that in the case of the $\theta^s_{\mu \nu}$
operator the transitions $\bar s s \to \bar u u + \bar d d$
are suppressed, while this is not the case for the $j^s_{5 \mu}$
operator --- the OZI rule works well in the first case and is violated in
the second. This circumstance is not surprising. In the meson nonets the
situation is similar: the OZI rule works well in the case of vector and
tensor mesons and is 100\% violated in pseudoscalar nonet. The latter
phenomenon was explained theoretically by attributing $\bar s s \to
\bar u u + \bar d d$ transitions in mesons to nonperturbative QCD effects
--- the instantons $^{52}$. In the field of instanton the transitions
$\bar s s \to \bar u u + \bar d d$ proceed in pseudoscalar and
longitudinal axial channels, but are forbidden in vector or tensor
channels.

It is plausible $^{51}$, that the same instanton mechanism results in large
$\bar s s \to \bar u u + \bar d d$ transition in $< p \mid j^s_{\mu 5}
 \mid p >$, but not in $< p \mid \theta^s_{\mu\nu} \mid p >$. If it is
indeed the case, then the two--component form of $s$--quark distribution
is expected $^{51}$
\be
   \begin{array}{cc}
   s_+ (x) + s_- (x) = (A_1/x)(1-x)^{\beta_s} + (A_2/x)(1-x)^{p'}~ , \\
   s_+ (x) - s_- (x) = B (1-x)^p ~ ,
   \end{array}
\ee
where $\beta_s \approx 5$ and $p' \approx p \approx 10-12$. It is easy to
see that for distributions (67) the inequality (23) is weakened to $\mid
\Delta s \mid \lsim 8 V_{2s}$ and (64), (65) are not in contradiction.

In the last year there was an impressive progress in experiment in the
problem in view. SMC presented $^{53}$ the results of new measurements of
$g_{1p}(x)$. The region of $x$, where the measurement were performed was
extended to $0.003 < x < 0.7.$ The SMC data are in a good agreement with
previous EMC and SLAC data in the domain of $x$ common for all experiments.
However, at $x < 0.015$ the SMC points lay higher, than the EMC extrapolation.
The SMC data are presented in Fig.2.
SMC found for $\Gamma_p$ at $\bar Q^2
= 10 ~GeV^2$
\be
   \Gamma_p ~ = ~ 0.136 ~\pm~ 0.011 ~\pm~ 0.011 ~ .
\ee
In the combined analysis of all data -- SLAC, EMC and SMC, it was obtained
$^{53}$
$$ ~~~~~~~~\Gamma_p ~ = ~ 0.142 ~\pm~ 0.008 ~\pm~ 0.011 ~ .
   ~~~~~~~~~~~~~~~~~~~~~~~~~~~~~~~~~~~~~~~~~~~~~~~~~~(68') $$

\noindent
\begin{tabular}{p{65mm}p{85mm}}
 & \underline{\bf Fig. 2}\\
 & The SMC data  $^{53}$ in comparison with the theory. The solid circles
(right-hand axes) show $xg_{1p}(x)$, the open boxes (left-hand axes)
show $\int^1_{x_m} g_{1p}(x)dx$. Only statistical errors are shown. The
solid square shows the SMC result for $\Gamma_p$ with statistical and
systematic error combined in quadrature. The solid and dashed lines are
theoretical predictions $^4$ and $^{11}$ correspondingly (see text).
The Ellis--Jaffe prediction (63) is marked by the cross.
\end{tabular}

If gluonic contribution to $\Gamma_p$ is neglected, then it follows:\\
from (68):
\be
   \Sigma ~ = ~ 0.26 ~\pm~0.14 ~ , ~~~~~~ \Delta s ~ = ~ -0.12 ~\pm~ 0.05 ~,
\ee
from (68$'$)
\be
  \Sigma ~ = ~ 0.31 ~\pm~ 0.13 ~ ,
  ~~~~~~~~ \Delta s ~ = ~ -0.10 ~\pm~ 0.05 ~.
\ee
Quite recently the E143 group at SLAC $^{54}$ presented the preliminary
result of
$g_{1p}(x)$ measurements at $0.029 < x < 0.8$ and $\bar Q^2 = 3 ~GeV^2$.
E143 found
\be
   \Gamma_p ~ = ~ 0.133 ~\pm~ 0.004 \pm 0.012 ~ , ~~~~~~~ \bar Q^2
    = 3 ~GeV^2 ~ .
\ee
(In obtaining (71) E143 group introduced a correction for $x < 0.029$
region contribution by extrapolation of their data. This extrapolation
curve however, lays below the direct SMC data points.)  E143 data (71)
result in
\be
   \Sigma ~ = ~ 0.32 ~\pm~ 0.12 ~, ~~~~~~~~~~ \Delta s ~ = ~ -0.10
{}~\pm~ 0.04 ~.
\ee
The neutron structure function $g_{1n}(x)$ was measured in two experiments:
by SMC at CERN in polarized muon scattering on polarized deuterium $^{55}$
and by E142 group at SLAC in polarized electron scattering on polarized
$^3He~ ^{56}$. The mean value of $\bar Q^2$ in the SMC experiment was
$\bar Q^2 = 4.6 ~GeV^2,$ in E142 experiment $\bar Q^2 = 2 ~GeV^2$. After
correction for $D$-wave admixture in the deuteron, SMC found
\be
   \Gamma_p + \Gamma_n ~ = ~ 0.050 \pm 0.044 \pm 0.033 ~,
   ~~~~~~~~~~~\bar Q^2 = 4.6 ~GeV^2 ~ .
\ee
In order to determine $\Gamma_n$  from (73) and $\Gamma_p$ measurements the
data must be taken at one common value $Q^2 = Q^2_0$. We will do such
recalculation, in the same way as in $^{57}$ -- accounting for perturbative
corrections given in (40) and twist 4 correction (56) and neglecting
$\Delta g$ contribution. As $Q^2_0$ we choose $Q^2_0 = 10.7 ~GeV^2$ .
We get
\be
   \frac{(\Gamma_p + \Gamma_n) \mid_{Q^2_0}}{(\Gamma_p + \Gamma_n)_{4.6
{}~GeV^2}}
    ~ = ~ 1.05
\ee
and the corresponding to (73) value at $Q^2_0$ is
\be
   \Gamma_p + \Gamma_n ~ = ~ 0.053 \pm 0.046 \pm 0.035 ~ , ~~~~~~~
   \bar Q^2_0 ~ = ~ 10.7 ~GeV^2 ~.
\ee
If the result of combined analysis (68$'$) is taken for $\Gamma_p$ then
\be
   \Gamma_n ~ = ~ -0.089 \pm 0.04 \pm 0.04
\ee
and
\be
   \Gamma_p - \Gamma_n ~ = ~ 0.231 \pm 0.045 \pm 0.045 ~, ~~~~~~~
    Q^2 ~ = ~ Q^2_0 ~ .
\ee
This experimental number may be compared with the Bjorken sum rule,
calculated with account of perturbative QCD corrections up to $\alpha^3_s$
$^{18}$
\be
   \left( \Gamma_p - \Gamma_n  \right)_{theor} ~ = ~ 0.186 \pm 0.003 ~,
    ~~~~~~Q^2 ~ = ~ Q^2_0 ~.
\ee
The two numbers agree within the experimental errors, which, unfortunately,
are rather large.

The E142 experiment has the advantage that the spins of two protons in
$^3He$ are compensated and the experiment (up to small correction)
gives directly $g_{1n}(x)$. The experiment was done at $0.03 < x < 0.6$.
Performing the extrapolation in the regions of small and large $x$ the
E142 group obtained
\be
   \Gamma_{1n} ~ = ~ -0.022 \pm 0.011 ~ , ~~~~~~~~~~~
   \bar Q^2 ~ = ~ 2 ~GeV^2 ~
\ee
Ellis and Karliner $^{57}$, using E142 data, but different extrapolation
obtained
instead $\Gamma_{1n} = -0.028$. The value $\Gamma_{1n}$ close to the latter
was obtained by SMC $^{58}$, when they use their data for extrapolation.
Transferring the E142 group data to the common value $Q^2 = Q^2_0$, we have
from (79)
\be
   \Gamma_n (Q^2_0) ~ = ~ -0.024 \pm 0.012 ~~(-0.030) ~.
\ee
(In the parenthesis the value, following from analysis $^{57, 58}$ is
given.) From (68$'$) and (80) it follows for the Bjorken sum rule
\be
   \Gamma_p - \Gamma_n ~ = ~ 0.166 \pm 0.021 ~~ (0.172)~, ~~~~~~~~
    Q^2 ~ = ~ Q^2_0 ~,
\ee
in agreement with theoretical value (78). The values of proton spin
projection carried by quarks can be also determined from E142 data (79).
The results are the following
\be
   \begin{array}{cc}
\Delta u = 0.88 \pm 0.05 ~, \Delta d = -0.37 \pm 0.05 ~,
\Sigma = 0.47 \pm 0.12 \\
    \Delta s = -0.04 \pm 0.05 ~ .
   \end{array}
\ee
(The difference between (79) and the value given by $^{57, 58}$ is included in
the error.) The results of all experiments, represented in terms of the
part of proton spin projection, carried by $u, d$ and $s$--quarks are
compatible,
although some spread in the results is also seen. (Particularely, the
difference between the EMC and E142 data.)

Different experiments were done at
different $\bar Q^2$ and, even more, each bin in $x$ corresponds to each own
$Q^2$. The problem arises, if the account of $Q^2$ dependence could change the
results. Experimentally, no $Q^2$ dependence in the asymmetry $A$ (7) was
observed, but the accuracy is not good enough --- not better than 10\% at high
$Q^2$ and 20\% at low $Q^2$. In ref.59 by solving the evolution equations
the problem was studied, how the fact, that each bin in $x$ is measured
at each own $Q^2$, can affect the final result. It was found
that this effect is small, less than the errors in the present experiments.

Till now when examining the experimental data the
higher twist corrections were disregarded, or accounted
basing on the calculations $^{41}$, where they are very small, much less than
experimental errors. However, as was explained in Sec.5, the results of
$^{41}$ are not convincing. Now I will discuss another way of determination
of $\Gamma_{p, n} (Q^2)$ $Q^2$ dependence, based on the connection of
$\Gamma_{p, n} (Q^2)$ with GDH sum rule.

Following $^{3, 6}$, introduce the functions
\be
   I_{p,n}(Q^2) ~ = ~ \int\limits^{\infty}_{Q^2 /2} \frac{d \nu}{\nu}
   G_{1, p, n} (\nu, Q^2) ~ .
\ee
Using (2), it is easy to demonstrate, that at large $Q^2$
\be
   I_{p,n} (Q^2)  ~ \approx ~ \frac{2m^2}{Q^2} \Gamma_{p,n} (Q^2) ~.
\ee
At $Q^2 = 0$, according to GDH sum rule, we have
\be
   I_{p,n} (0) ~ = ~ - \frac{1}{4} \kappa^2_{p,n} ~.
\ee
\vspace{3cm}

\noindent
\begin{tabular}{p{60mm}p{80mm}}
 & \underline{\bf Fig. 3}\\
 & The connection of GDH sum rule with sum rules at high $Q^2$ - the
qualitative dependence of $I_p(Q^2), I_n(Q^2)$ and $I_p(Q^2) -
I_n(Q^2)$ on $Q^2$.
\end{tabular}
\vspace{3cm}

\noindent
The $Q^2$ dependence of $I_{p,n} (Q^2)$, and the difference $I_p (Q^2) --
I_n (Q^2)$
is plotted qualitatively in Fig. 3. The case of $I_p$  is
especially interesting. At large $Q^2 ~ I_p(Q^2)$ is positive, but at
$Q^2 = 0$ it is negative and large, $I_p(0) = -0.72$. This means that in the
region $Q^2 \lsim 1 ~GeV^2, ~ \Gamma_p(Q^2)$ varies strongly and changes
the sign. Such a behavior cannot be achieved by smooth extrapolation of
perturbative QCD corrections and is an indication of large nonperturbative
effects.

In ref.6 the vector dominance based model was suggested, which interpolates
$I_{p, n}(Q^2)$ at intermediate $Q^2$, has the correct asymptotics (84) at
large $Q^2$ and satisties the GDH sum rule (85) at $Q^2 = 0$. In constructing
such model it must be taken into account, that the contribution of
baryonic resonances are important in $I_{p,n}(Q^2)$ at low $Q^2$ $^{60, 61}$ .
Therefore, the model for $I_{p,n}(Q^2)$ has the form
\be
   I_{p,n} (Q^2) ~ = ~ I_{p, n}^{res} (Q^2)  +  I_{p, n}' (Q^2) ~ ,
\ee
where $I_{p,n}^{res}(Q^2)$ is the contribution of baryonic resonances,
known from the analysis of pion electroproduction experiments (up to mass
$W = 1.8 ~GeV$) $^{62}$. $I_{p, n}' (Q^2)$  is defined by
\be
   I'_{p,n} (Q^2) = 2m^2 \Gamma^{as}_{p,n}
   \left[  \frac{1}{Q^2 + \mu^2} - \frac{c_{p,n} \mu^2}{(Q^2 + \mu^2)^2}
   \right] ~ ,
\ee
\be
   c_{p,n} = 1 + \frac{1}{2} \frac{\mu^2}{m^2} \frac{1}{\Gamma^{as}_{p,n}}
   \left[  \frac{1}{4} \kappa^2_{p,n} + I^{res}_{p,n}(0)  \right]  ~,
\ee
where $\mu^2$ is the vector meson mass, $\mu^2 = 0.6 ~GeV^2.$
$\Gamma^{as}_{p,n}$ means the value $\Gamma_{p,n}$ at large $Q^2$ with higher
twist terms excluded. For our purposes --- to determine the $Q^2$ power
corrections we may consider $\Gamma^{as}_{p,n}$ as $\Gamma_{p,n}$  at
$Q^2 \sim 10 ~GeV^2$. The second term in (87) in the framework of VDM
corresponds to the case when  in forward virtual $\gamma$
Compton scattering amplitude both $\gamma$  interact with
 the nucleon through vector mesons. The first term in (87) corresponds
to the case, when one $\gamma$  interacts  through vector
meson and the other one directly. It is easy to see that in the model the
asymptotic behavior of $I_{p,n}(Q^2)$ and GDH sum rule is fullfilled. From
the experimental data on pion photoproduction, the known values of $\kappa
_{p,n}$ and experimental data of $\Gamma^{as}_{p,n}$ it was found $^{61}$
\be
   c_p = 0.43 \pm 0.10 ~ , ~~~~~~~~~~~ c_n = 0.0
   \begin{array}{l}
    +0.3 \\ -1.2
   \end{array}
\ee
The large error in $c_n$ results from large uncertainty in $\Gamma_n$
in E142 experiment.

The resonance contribution is absent (or small) in the region of $Q^2$
and $x$, where the existing experiments were performed. Then the power
corrections came from $I'_{p,n}$ in (86) and can be calculated using (87)
- (89). The results are
\be
   \begin{tabular}{lllll}
   $Q^2 (GeV^2)$                    & 2    & 3    & 4.6  & 10.5 \\
   $\Gamma^{as}_p/\Gamma^{exper}_p$ & 1.44 & 1.29 & 1.19 & 1.08  \\
   $\Gamma^{as}_n/\Gamma^{exper}_n~~~$ & 1.30~ & 1.20~ & 1.13~ & 1.06~
   \end{tabular}
\ee
The results for $\Gamma_{p,n}, \Sigma$ and $\Delta s$ arising after
introduction of power corrections (90) are given in the  Table I.
(The SMC $\Gamma_n$ data are not included, because of large uncertainty
in $\Sigma$ and $\Delta s$).
\vspace{2mm}

\underline{Table I}

\begin{tabular}{|l|c|c|c|c|} \hline
& EMC(p) & SMC(p) &  E143(p) & E142(n)  \\  \hline
$\Gamma_{p,n}$ & $0.137 \pm 0.018$ & $0.147 \pm 0.016$ & $0.172 \pm 0.013$ &
       $-0.029 \pm 0.015$ \\  \hline
$\Sigma$       & $0.28 \pm 0.17$   & $0.37 \pm 0.14$   &
       $0.68 \pm 0.12$ & $0.46 \pm 0.13$ \\  \hline
$\Delta s$     & $-0.10 \pm 0.06$ & $-0.08 \pm 0.05$ &
       $+0.02 \pm 0.04$ & $-0.04 \pm 0.05$ \\ \hline
\end{tabular}
\vspace{2mm}

All the data after introduction of power corrections are consistent with
$\Sigma \approx 0.4-0.5$ and $\Delta s \approx -0.02-0.04$ --- the
values close to naive expectations --- and no large gluonic term is needed.
The contribution of gluons was disregarded in the calculation of $\Delta u,
\Delta d, \Delta s$ and $\Delta \Sigma$, the results of which are presented in
(69), (70), (72), (82) as well as in Table I. As was mentioned above, if
$\Delta g (10 ~GeV^2) = 0.5$ the account of gluons would increase $\Sigma$
and $\Delta s$ by 0.07 and 0.02 correspondinly and twice as large if
$\Delta g (10 ~GeV^2) = 1$, what, may be, is also acceptable.

Till now I discussed mostly the integrals of $g_1(x)$ over $x$, which are
known better theoretically than the $x$-dependence of $g_1(x)$. There are few
theoretical models for $g_1(x)$ but, unfortunately, the successful description
of the
data was achieved by the price of use some experimental information as input.
Fig.2 displays the theoretical curve for $g_{1p}(x)$, given by Brodsky $^{4}$,
on the base of matching the Regge behavior at small $x$ with quark counting
at large $x$. An essential point was that the matching was done separately
for $q_+(x)$ and $q_-(x) (q = u, d, s).$ The values $\Delta u, \Delta d,
 \Delta s $ from EMC experiment are used as input. The same figure shows also
$g_{1p}(x)$ found in the QCD sum rule approach $^{11}$. In this calculation
no experimental input is used: the only parameter, which enters the
calculation is the fundamental QCD parameter - the quark condensate
$\alpha_s < 0 \mid \bar{\psi} \psi \mid 0 >^2$. Unfortunately, the domain
of $x$, where the results $^{11}$ are reliable (this domain was determined in
the calculation) is rather narrow, $0.5 \le x \ge 0.8$.
\vspace{5mm}

\noindent
{\large\bf 9. Suggestions for Future Experiments}
\vspace{3mm}

1. \underline{\it The experimental study of the sum rule $Q^2$ dependence.}
As was discussed in the previous Section, the part of the deviation from the
theoretical expectations, as well as the difference  between the data,
obtained at different $Q^2$, may be attributed to high twist effects. Also,
the role of $Q^2$ dependent $\alpha_s$ corrections is not negligible in the
problem in view. For this reason the precise measurements of the $Q^2$
dependence of the sum rules would be very desirable. Such measurements are
expected in the near future at SLAC (see $^{63, 64}$), CERN and DESY
(HERMES Collaboration). Especially interesting would be the direct experimental
test of the GDH sum rule and the study of $Q^2$ dependence of $I_{p,n}(Q^2)$
(83) in the domain $0 < Q^2 \lsim 2 GeV^2$. Up to now the GDH sum rule was
checked in indirect way, using the data on the photoproduction of baryonic
resonances. It was found $^{65, 66}$, that if the GDH sum rule is decomposed
into isoscalar and isovector components, then the isovector--isovector sum
rule is well satisfied, isoscalar--isoscalar contribution is small. However,
in the isovector--isoscalar sum rule the contribution of high mass
resonances is essential, what deteriorates the accuracy of the results.
The study of $I_{p,n}(Q^2) ~ Q^2$ dependence would be important for
checking of various models. Particularely, in the model $^{61}$ it is
expected, that $I_p(Q^2)$ is crossing zero at $Q^2 \approx 0.5 ~GeV^2$.

2. \underline{\it The search for $\Delta g$.} The straightforward way to
determine $g_+(x) - g_-(x)$ is by mearuring two jets in the virtual photon
fragmentation region in the polarized $\mu(e)$-- proton deep inelastic
scattering $^{20}$. Such events arise from $\gamma - g \to \bar q q$
collisions and may be separated from other two jet events by their known
$k^2_{\perp}$ dependence $^{20}$. When the jets are well separated in
$k^2_{\perp}$, there are no infrared problems and the cross section is
predicted uniqully in terms of $g_+(x) - g_-(x)$.

For discussion of other possibilities see the review $^{32}$ and
references herein.

3. \underline{\it The measurements of $\Delta s$.} Besides of $\Delta s$
determination in future precise polarized deep inelastic $\mu(e)$--nucleon
scattering, $\Delta s$ can also be measured in elastic $\nu p$ scattering
by separating axial formfactor contribution at zero momentum transfer $^{67}$.
The axial formfactor arises from $Z$--boson exchange and at $Q^2 = 0$
is proportional to the matrix element
\be
   < p \mid \bar u \gamma_{\mu} \gamma_5 u - \bar d \gamma_{\mu} \gamma_5 d
   - \bar s \gamma_{\mu} \gamma_5 s \mid p > = - 2 s_{\mu} m (\Delta u -
   \Delta d - \Delta s ) ~.
\ee
Such an experiment gives an additional equation for determination of
$\Delta u, \Delta d, \Delta s $. The existing data $^{68}$ indicate
on large $\Delta s$
\be
   \Delta s ~ = ~ -0.15 ~ \pm ~ 0.08 ~ .
\ee
The determination of axial formfactor $G_A(0)$ from the data must be done
with some care. The experiment is performed at $Q^2 \ne 0$ and the
extrapolation to $Q^2 = 0$ proceeds, assuming the dipole formula for the
formfactor,
\be
   G_A (Q^2) ~ = ~ \frac{G_A (0)}{[1 + Q^2 / M^2_A]^2} ~ .
\ee
However, due to chirality conservation, (93) cannot be correct in the whole
domain of $Q^2$. This follows from the requirement that at large $Q^2$
the axial formfactor must coincide with vector one $^{69}$. Therefore
a deviation from the dipole fit (93) is expected at intermediate $Q^2.$
This expectation is in accord with the discussed in Sec.8 nonperturbative
effects (instantons) in flavour singlet axial formfactor. Surely, the
observation of the deviation from the dipole fit in singlet axial would be
extremingly important for our understanding of nonperturbative QCD.

The measurements of $s$--quark distribution in nonpolarized nucleon would
be also helpful in understanding the origin of the strange sea. The
$s$--quark distribution obtained from the data $^{70}$
on charm production in
$\nu N$ scattering shows a more steep $1 - x$ dependence
$$
  x s (x) ~ \sim ~ (1 - x)^{\beta} ~ , ~~~~~~~~ \beta \approx 6.5 ~ ,
$$
then the quark counting rule expectation $\beta \approx 5$.
This may be an indication in favour of nonperturbative (instanton)
component (67) in $s$--quark distribution, but more precise data are needed.

If $\Delta s$ is large and negative, then a surplus of jets with leading
strange
particles can be expected in the direction opposite to the proton spin. This
also could be tested.

4. \underline{\it The check of the OZI rule in axial mesons.} A connected
problem is the determination if OZI rule holds in the nonet of axial mesons.
If $\Delta s$ is large and nonperturbative effects are important in singlet
axial channel, then it is expected $^{48}$ a noticeable OZI rule violation
in the nonet of axial $1^{++}$ mesons. There are indications, that it is
indeed,
what happened $^{71}$.
\vspace{5mm}


\noindent
{\large\bf 10. The Structure Function $g_2 (x)$}
\vspace{3mm}


Since there are very good reviews $^{72, 32}$ on this subject and almost
nothing new happened after their appearance, I will indicate here the basic
points only. The structure function $g_2 (x)$  can be measured in deep
inelastic
$\mu(e)$--nucleon scattering, when $\mu(e)$  is polarized longitudinally and
nucleon transverse to the beam direction. In this case the contributions of
$g_1$ and $g_2$
to spin asymmetry (measured, e.g. by changing the direction of nucleon
spin to the opposite) are of the same order, but suppressed by the factor
$\sim 1/Q$ in comparison with the $g_1$ contribution in longitudinally
polarized scattering. In the cases of other directions of nucleon spin $g_2$
contribution is smaller, than $g_1$ by a factor $\sim 1/Q$. This fact
demonstrates that $g_2$ may be considered as a twist-3 structure function.

For free particle $g_2 \equiv 0$. This is an indication, that $g_2$ cannot be
described in terms of parton distributions, because in the parton model we
consider deep inelastic scattering, as proceeding on free partons. The other
argument in the same direction can also be presented. Let us suppose,
following Feynman $^{73}$ that the transversely polarized nucleon in the
infinite momentum frame is described by parton densities $t^{\pm}_j(x)$
with polarizations $\pm 1$ upon nucleon polarization direction. Then, it can be
shown $^3$, that
\be
    g_1(x) + g_2(x) = \sum e^2_j \left( \frac{\mu_j}{m} +
   \frac{\overline{p^2_{\perp j}}} {m Q}  \right)   \frac{1}{2x}
   \left[  t^+_j (x) - t^-_j (x)  \right]  ~ .
\ee
Here $\mu_j, e_j$ and $\overline{p^2_{\perp j}}$ are the mass, charge and
mean square of the transverse momentum of parton $j$. At  large $Q$, since
$\mu_j$ are the current masses, the r.h.s. of (94) is small. But the
integral over $x$ of the l.h.s. is not small, according to sum rules, so a
contradiction arises.  The quark-gluon interaction and transverse momentum
quark distribution in nucleon are of importance in $g_2(x)$ determination.
Both of them are beyond the parton model. For this reason our theoretical
knowledge of $g_2(x)$  is very scarce now.

In OPE on the light cone $g_2(x)$ can be decomposed into two sets of operators
$^{74}$. The first is twist-2 operators, the same, which appear in $g_1(x)$
decomposition, the second is twist-3 operators. The first set contribution can
be expressed through $g_1(x)$  and the relation can be obtained $^{74}$
\be
   g_2 (x, Q^2) = g_2^{WW} (x, Q^2) + \bar g_2 (x, Q^2) ~ ,
   ~~g_2^{WW}(x, Q^2) = -g_1 (x, Q^2) +
   \int\limits^1_x \frac{dz}{z} g_1 (z, Q^2) ~ ,
\ee
where $g^{WW}_2$ and $\bar g_2$ correspond to twist-2 and 3. There is no
reason to believe, that the most interesting piece --- $g_2 (x, Q^2)$,
determined by quark-gluon interaction, is much less than $g^{WW}_2(x, Q^2)$.
Therefore, it is unclear, if decomposition (95) is helpful.

It is very important to check experimentally the Burkhardt-Cottingham (BC)
sum rule (30). At such investigation it would be interesting to study $Q^2$
dependence of the sum rule: as was argued in Sec.4 one may expect that BC
sum rule is valid only in low twist order and at $Q^2 \sim 1 - 2 ~GeV^2$
the violation of BC sum rule could be seen.
\vspace{5mm}

\noindent
{\large\bf 11. Chirality Violating Structure Function $h_1(x)$}
\vspace{3mm}

As is well known, all structure functions of the twist two -
$F_1(x), ~F_2(x), ~g_1(x)$ which are measured in the deep-inelastic
lepton-nucleon scattering conserve
chirality. Ralston and Soper $^{75}$ first
demonstrated that besides these structure functions, there exists the
twist-two chirality violating nucleon structure function $h_1(x)$.
This structure function does not manifest itself in the deep inelastic
lepton-hadron scattering, but can be measured in the Drell-Yan process with
both beam and target transversally polarized. The reason of this circumstance
is the following. The cross section of the deep inelastic electron(muon)-hadron
scattering is proportional to the imaginary part of the forward virtual
photon-hadron scattering amplitude. At high photon virtuality the quark
Compton amplitude dominates, where the photon is absorbed and emitted by the
same quark (Fig.4a) and the conservation of chirality is evident.
 The  cross section of the Drell-Yan process can be represented as an
imaginary part of the diagram, Fig.4b. Here virtual photons interact with
different quarks and it is possible, as  is  shown  in
Fig.4b, that chirality violating amplitude in Drell-Yan processes is not
suppressed at high $Q^2$ in comparison with chirality conserving ones,
and consequently, corresponds to twist two. However, this amplitude,
corresponding to target spin flip, has no parton interpretation in terms
of quark distritutions in the helicity basis and, as was shown by Jaffe
and Ji $^{76}$ can be only represented as an element of the quark-quark
density matrix in this basis.
\vspace{5mm}

\noindent
\begin{tabular}{p{75mm}p{75mm}}
 & \underline{\bf Fig. 4}\\
 & a) Deep inelastic lepton-hadron scattering, the quark chiralities are
conserved. Solid lines are quarks, wavy lines are virtual photons, $R(L)$
denote right (left) chirality of quarks; b) Drell-Yan process with
chirality of quarks flipped.
\end{tabular}
\vspace{5mm}

\noindent
However, $h_1(x)$ can be interpeted $^{76}$  as a difference of quark
densities with the eigenvalues $+1/2$ and $-1/2$ of the transverse
Paili-Lubanski spin operator $\hat s_{\perp} \gamma_5$ in the transversely
polarized proton. It this basis $h_1(x)$ can be described in terms of
standard parton language.

Until now there are no experimental data on the chirality violating
nucleon structure function $h_1(x)$. Besides $^{75, 76}$ the theoretical
study of this structure function have been also performed by Artru and
Mekhfi $^{77}$. The role of $h_1(x)$ in factorization of a general hard
process with polarized particles was investigated by Collins $^{78}$.
The first attempt to calculate $h_1(x)$ was carried out by Jaffe and
Ji $^{76}$ by means of the bag model.

The proton structure function $h_1(x)$ can be defined in the light cone
formalism as follows $^{76}$
\be
   \begin{array}{cc}
i \int \frac{d \lambda}{2 \pi} e^{i \lambda x} < p, s \mid \bar{\psi}
(0) \sigma_{\mu \nu} \gamma_5 \psi (\lambda n) \mid p, s > =
2 [ h_1 (x, Q^2)(s_{\perp \mu} p_{\nu} - s_{\perp \nu} p_{\mu})
+ \\
   + h_L (x, Q^2) m^2 (p_{\mu} n_{\nu} - p_{\nu} n_{\mu}) (s n) +
   h_3 (x, Q^2) m^2 (s_{\perp \mu} n_{\nu} - s_{\perp \nu} n_{\mu}) ] ~.
   \end{array}
\ee
Here $n$ is a light cone vector of dimension (mass)$^{-1}, n^2 = 0,
n^+ = 0, pn = 1, p$ and $s$ are the proton momentum and spin vectors,
$p^2 = m^2, s^2 = -1, ps = 0$ and $s = (sn)p + (sp)n + s_{\perp},
h_{L}(x, Q^2)$ and $h_3 (x, Q^2)$ are twist-3 and 4 structure functions.
For comparison in the same light cone notation the standard structure
function $F_1(x, Q^2)$ is given by
\be
   \int \frac{d \lambda}{2 \pi} e^{i \lambda x} < p, s \mid \bar{\psi}
   (0) \gamma_{\mu }  \psi (\lambda n) \mid p, s > =
   4 [F_1 (x, Q^2) p_{\mu} + M^2 f_4 (x, Q^2) n_{\mu} ]
\ee
(Eq. (96), (97) are written for one flavour).

Basing on the definition (96) an inequality was proved in $^{76}$
\be
   q (x) ~ \ge ~ h_1 ^q (x) ~ ,
\ee
which holds for each flavour $q = u, d, s$. (Here $h_1 ^q (x)$ is the
flavour $q$ contribution to $h_1 (x)$.)  It was also suggested $^{76}$,
that $ \mid h_1 ^q (x) \mid  > \mid g_1 ^q (x) \mid $.

$h_1 (x)$ can be also represented through a $T$-product of currents $^{79}$
\be
   T_{\mu}(p, q, s) = i \int d^4 x e^{iqx} < p, s \mid (1/2) T
   \{ j_{\mu 5} (x), j (0) + j (x), j_{\mu 5} (0) \} \mid p, s > ~ ,
\ee
where $j_{\mu 5} (x)$ and $j(x)$ are axial and scalar currents.

The general form of $T_{\mu}(p, q, s)$ is
$$  T_{\mu}(p, q, s) =  \left( s_{\mu} - \frac{q s}{q^2} q_{\mu} \right)
    \tilde h_1 (x, Q^2) + \left( p_{\mu} - \frac{\nu q_{\mu}}{q^2} \right)
    (q s ) l_1 (x, Q^2) ~ +
$$
\be
   + ~ \varepsilon_{\mu \nu \lambda \sigma} p_{\nu} q_{\lambda}s_{\sigma}
   (q s) l_2 (x,q^2)
\ee
(only spin--dependent terms are retained). It can be proved $^{79}$, that
\be
   h_1 (x, Q^2) ~ = ~ -\frac{1}{\pi} Im \tilde h_1 (x, Q^2) ~ .
\ee
As is clear from (96) or (99) $h_1(x)$ indeed violates chirality.
It is worth mentioning that instead of axial and scalar currents in (99)
the combination of vector and pseudoscalar currents $j_{\mu} (x) j_s (0)
- j_s (x) j_{\mu} (0)$ can also be used (notice the minus sign in the
crossing term).

The structure function $h_1(x)$ can be measured in the production of
$\mu^+ \mu^-$ or $e^+ e^-$ pairs in $pp$ (or $p \bar p$) collisions,
when both protons (or $p$ and $\bar p$) are polarized transversally to
the beam direction. Introduce the coordinate system, where the beam direction
is along $z$, the proton spin direction---along $x$ and the momentum of the
lepton pair is charachterized by polar and azimuthal angles $\theta$ and
$\varphi$ (in the lepton pair c.m.s.). Then the asymmetry is given by $^{75,
80}$
\be
  A(x,y) = \frac{\sigma_+ (x,y) - \sigma_- (x,y)} {\sigma_+
            (x,y) - \sigma_- (x,y)}   = \frac{\sin \theta \cos 2 \varphi}
  {(1 + \cos^2 \theta)} ~\frac{\sum_a e^2_a h_1^a (x)  h^{\bar a}_1 (y)}
  {\sum_a e^2_a q^a (x)  q^{\bar a}_1 (y)}
\ee

Here $\sigma_+ (x, y)$ and $\sigma_-
(x, y)$ are the cross sections of lepton pair production, where the spin of
one of protons is along or opposite to spin of the other (or $\bar p$).

In (102) $x$ and $y$ are the part of the momenta, carried by quarks and
antiquarks, annihilating into virtual photon with momentum $q$
\be
   \begin{array}{rr}
       q ~ = ~ x p_1 + y p_2 + q_{\perp} ~ ,\\
       M^2 ~ = ~ q^2 ~ =  ~ x y s -
       \mbox{\boldmath$q$}^2_{\perp}  ~ ,\\
   \end{array}
\ee
where $M$ is the mass of leptonic pair. $h^a_1$ and $h^{\bar a}_1$  are
the structure functions $h_1$ for quarks and antiquarks in the target and
projectile  correspondingly, $q^a$ and $q^{\bar a}$ are the quark and antiquark
distributions. In $p \bar p$ collisions $h^a_1 = h^{\bar a}_1, q^a = q^{\bar
a}$.

Since $h_1(x)$ violates chirality one may expect that in QCD it can be
expressed in terms of chirality violating fundamental parameters of the
theory, the simplest of which (of lowest dimension) is the quark condensate.
Basing on this idea $h_1(x)$ calculation was performed in $^{79}$. The
general method $^{81, 82}$ of structure function calculations in the QCD
sum rule approach was used. In this method the structure function at
intermediate $x$ can be found. The main difference in comparison with $^{82}$
is that in the case of $h_1(x)$ determination the chirality violating
structure is studied. As a result $h_1(x)$ was found to be proportional to the
quark condensate $< 0 \mid \bar u u \mid 0 >$  with the correction term in
OPE proportional to the mixed quark--gluon condensate $g < 0 \mid \bar u
\sigma_{\mu \nu} G^n_{\mu \nu} \lambda^n u \mid 0 >$. It was obtained that
for proton $h^u_1 (x) \gg h^d_1 (x)$, and as a consequence $h_1 (x) \approx
(4/9) h^u_1 (x)$. The calculations in $^{79}$ are valid at $0.3 < x < 0.55$.
The extrapolation in the region of small $x$ can be performed using Regge
behavior $h_1 (x) \sim x^{-\alpha_{a_1}}$,
the extrapolation in the region
of large $x$, using the inequality (98). The final result of calculation of
$h_1^u (x)$  with the extrapolations is shown in Fig.5. \\
\vspace{10mm}

\noindent
\begin{tabular}{p{70mm}p{80mm}}
 & \underline{\bf Fig. 5}\\
 & The $u$-quark contribution to the proton structure function $h_1(x)$
based on QCD sum rule calculation $^{79}$ at intermediate $x,
{}~0.3 < x < 0.55$. At $x < 0.3$ an extrapolation according to Regge behavior
was performed, at $x > 0.55$ (dashed line) the saturation of inequality (98)
was assumed.
\end{tabular}
\vspace{10mm}

\noindent
It is expected that
the accuracy of $h^u_1 (x)$ determination is about 30\% at $x \approx 0.4$
and about 50\% at $x = 0.6$. The inequality $h^u_1 (x) > g^u_1 (x)$
suggested in $^{76}$ was confirmed. Numerically $h_1(x)$ is rather large, what
gives a good chance for its experimental study.

It is difficult to produce intense high energy polarized antiproton beam
($\bar p p$ collisions are preferable for the study of $h_1$ structure
function, since the same $h^a_1$ enters (102) for target and projectile).
For this reason there are suggestions of experiments, where only target proton
is polarized, but also the polarization of the final particle is measured
$^{83-86}$
\vspace{5mm}


\noindent
{\large\bf 12. ~C o n c l u s i o n s}
\vspace{3mm}


In the last year the situation, connected with the so called \lq\lq proton
spin crises\rq\rq ~became less dramatic, than it was before, after the
appearance of the first EMC data. Although there is no contradiction between
the new and the old data, the world averages moved towards
theoretical expectations. The simplest theoretical prediction---the
Ellis--Jaffe
sum rule value moved slightly towards  experimental number
after appearance of new data on hyperon $\beta$--decay. As a result, the gap
between the experimental value of $\Gamma_p$ and Ellis--Jaffe sum prediction
shrinked. The new neutron data are more or less in agreement with theoretical
expectations and there is no  contradiction with the Bjorken sum rule.

On the other side the theoretical understanding of the nucleon spin content
increased essentially: we understand now that Ellis--Jaffe sum rule is not the
last word in the problem, we have more knowledge about the role of gluons and
strange  quarks. Since the gap between the theoretical expectation given
by Ellis--Jaffe sum rule and world averaged experimental value became narrower,
a possibility appeared to share this difference between the sourses,
nonaccounted in Ellis--Jaffe sum rule (strange sea, gluons, higher twist terms)
in more or less conservative way. For example, such a scenario is not in
contradiction with experimental data: at $Q^2 = 10 ~GeV^2~ \Delta g \approx
0.5 - 1.0, ~ \Delta s \approx -0.02 - 0.05, ~\Sigma \approx 0.4 - 0.5$ and
higher twist corrections comprising $\sim 4\%$ (the latter two times less
than in Table 1).

Evidently, these numbers would surprise nobody and would not require any
revolutionary changes in our understanding of nucleon structure. The problem
is:
if this or a similar scenario is realized in the nature. Future, more precised
experiments on deep inelastic $\mu(e)$ nucleon scattering will help us to
obtain the answer this question. I emphasize, that these experiments, even very
precised, are not enough for the final solution of the problem: an independent
experiments, measuring the part of the nucleon spin carried by strange quarks
and gluons are needed. A new era in the investigation of the nucleun spin
structure will open, when the measurements with the transversally (relative
to the beam) polarized nucleon will start. May be new surpises are waiting us
here.

I am grateful to J.Lichtenstadt for the discussion of experimental results and
for the information about new data
and to M.Karliner for helpful discussion of new theoretical developments.

This work was supported in part by International Science Foundation Grant
 M9H000 and by International Association for the Promotion of Cooperation
with Scientists from the Independent States of the Former Soviet Union Grant
INTAS--93--283.

\newpage

\noindent
{\large\bf ~R e f e r e n c e s}
\vspace{5mm}


\noindent
1.  J.Ashman et al., {\it Nucl Phys.} {\bf B238} (1989) 1.\\
2.  M.J.Alguard et al., {\it Phys. Rev. Lett.} {\bf 37} (1976) 1258, 1261.

    G.Baum et al., {\it Phys. Rev. Lett.} {\bf 51} (1983) 1135.\\
3.  B.L.Iof\/fe, V.A.Khoze, L.N.Lipatov,  {\it Hard Processes, v.1}

    (North Holland, Amsterdam, 1984), Sec. 2.10, 4.4., 6.4.6.\\
4.  S.J.Brodsky, in: {\it Lecture at SLAC Summer Institute on
    Particle Physics,

    July 26 - August 6, 1993} (Preprint SLAC-PUB-6450, 1994).\\
5.  G.Preparata, J.Sof\/fer, {\it Phys. Rev. Lett.} {\bf 61} (1988) 1167.\\
6.  M.Anselmino, B.L.Iof\/fe, E.Leader, {\it Sov. J. Nucl. Phys.}
{\bf 49} (1989) 136.\\
7.  B.L.Iof\/fe, {\it Phys. Lett.} {\bf 30B} (1969) 123.\\
8.  J.D.Bjorken, {\it Phys. Rev.} {\bf 148} (1966) 1467.\\
9.  J.Ellis, R.L.Jaffe, {\it Phys. Rev.} {\bf D9} (1974) 1444, {\bf D10} (1974)
    1669.\\
10. H.Burkhardt, W.H.Cottingham, {\it Ann. of Phys.} {\bf 56} (1970) 453.\\
11. V.M.Belyaev, B.L.Ioffe, {\it Int. Journ. of Mod. Phys.} {\bf A6}
    (1991) 1533.\\
12. S.B.Gerasimov, {\it Yad. Fiz. } {\bf 2} (1965) 598.

    S.D.Drell, A.C.Hearn, {\it Phys. Rev. Lett} {\bf 16} (1966) 908.\\
13. M.A.Ahmed, G.G.Ross, {\it Nucl. Phys.} {\bf B111} (1976) 441.\\
14. J.Kodaira et al., {\it Phys. Rev.} {\bf D20} (1979) 627;

    {\it Nucl. Phys.} {\bf B159} (1979) 99, {\bf B165} (1980) 129.\\
15. C.S.Lam, Bing-An Li, {\it Phys. Rev.} {\bf D25} (1982) 683.\\
16. A.V.Efremov, O.V.Teryaev, {\it Dubna Preprint JINR-E2-88-287} (1988).\\
17. G.Altarelli, G.G.Ross, {\it Phys. Lett. } {\bf B 212} (1988) 391.\\
18. S.A.Larin, J.A.M.Vermaseren, {Phys. Lett.} {B259} (1991) 345.\\
19. S.A.Larin, {\it Preprint CERN-TH} 7208/94 (1994).\\
20. R.D.Carlitz, J.C.Collins, A.H.Mueller, {\it Phys. Lett.} {\bf B214}
    (1988) 229.\\
21. R.L.Jaffe, A.Manohar, {\it Nucl. Phys.} {\bf B337} (1990) 509.\\
22. S.Forte, {\it Phys. Lett.} {\bf B224} (1989) 189;
    {\it Nucl. Phys.} {\bf B331} (1990) 1.\\
23. G.Altararelli, W.J.Stirling, {\it Particle World} {\bf 1} (1989) 40.\\
24. G.Veneziano, {\it Mod. Phys. Lett.} {\bf A4} (1989) 1605.

    G.M.Shore, G.Veneziano, {\it Phys. Lett} {\bf B244} (1990) 75,
     {\it Nucl. Phys.}  {\bf B381}

     (1992) 23. \\
25. G.Altarelli, B.Lampe, {\it Z. Phys.} {\bf C47} (1990) 315.\\
26. G.T.Bodwin, J.Qiu, {\it Phys. Rev. } {\bf D41} (1990) 2755.\\
27. L.Mankiewicz, A.Sch\"afer, {\it Phys. Lett.} {\bf B242} (1990) 455;

    L.Mankiewicz, {\it Phys. Rev. } {\bf D43} (1991) 64.\\
28. T.P.Cheng, L.F.Li, {\it Phys. Rev. Lett. } {\bf 62} (1989) 1441,

    Carnegie-Mellon Univ. Preprint CMU-HEP-90-2 (1990).\\
29. A.V.Manohar, {\it Phys. Rev. Lett. } {\bf 66} (1991) 289.\\
30. W.Vogelsang, {\it Z. Phys.} {\bf C50} (1991) 275.\\

\newpage
\noindent
31. S.D.Bass, B.L.Iof\/fe, N.N.Nikolaev, A.W.Thomas, {\it J. Moscow Phys.
    Soc.} {\bf 1}

    (1991) 317.\\
32. E.Reya, in: {\it Proc. of Intern. Workshop "QCD-20 Years Later", Aachen,
    1992 }

    (World Scientific, 1993).\\
33. S.J.Brodsky, I.Schmidt, {\it Phys. Lett.}  {\bf B34} (1990) 144.\\
34. M.Gl\"uck, E.Reya, A.Vogt, {\it Z. Phys.} {\bf C53} (1992) 127.\\
35. M.Gl\"uck, E.Reya, {\it Phys. Lett.} {\bf B270} (1991) 65.\\
36. J.Ellis, M.Karliner, C.T.Sachraida, {\it Phys. Lett. } {\bf B231}
    (1989) 497.\\
37. S.D.Bass, N.N.Nikolaev, A.W.Thomas, {\it Preprint Univ. of Adelaide

     ADP-133-T-80 } (1990).\\
38. E.V.Shuryak, A.I.Vainstein, {\it Nucl. Phys.} {\bf B201} (1982) 141.\\
39. B.L.Ioffe, A.V.Smilga, {\it Pisma v ZhETF} {\bf 37} (1983) 250;
    {\it Nucl.Phys.} {\bf B232}

    (1984) 109.\\
40. I.I.Balitsky, A.V.Yung, {\it Phys. Lett.} {\bf B129} (1983) 328.\\
41. I.I.Balitsky, V.M.Braun, A.V.Kolesnichenko, {\it Phys. Lett.}
    {\bf B242} (1990) 245

    {\it Errata} {\bf B318} (1993) 648.\\
42. Particle Data Group, M.Aguilar-Benitez et al., {\it Phys. Rev.}
    {\bf D45}, Part 2 (1992).\\
43. S.Y.Hsueh et al., {\it Phys. Rev.} {\bf D38} (1988) 2056. \\
44. H.J.Lipkin, {\it Phys. Lett.} {\bf B256} (1991) 284.\\
45. V.M.Belyaev, B.L.Ioffe, Ya.I.Kogan, {\it Phys. Lett} {\bf 151B}
    (1985)  290.\\
46. S.J.Brodsky, J.Ellis, M.Karliner, {\it Phys. Lett} {\bf B206}
    (1988) 309.

    J.Ellis, M.Karliner, {\it Phys. Lett. } {\bf B213} (1988) 73.\\
47. Z.Ryzak, {\it Phys.Lett.} {\bf B217} (1989) 325.

    V.Bernard, U.G.Meissner, ibid {\bf B223} (1989) 439.\\
48. B.L.Ioffe, A.Yu.Khodjamirian, {\it Yad. Fiz.} {\bf 55} (1992) 3045.\\
49. H.Fritzsch, {\it Phys. Lett} {\bf B329} (1989) 122, ibid {\bf B242}
    (1990) 451, ibid {\bf B256}

    (1991) 75.

    A.V.Efremov, J.Soffer, N.A.T\"ornqvist, {\it Phys. Rev. Lett.}
    {\bf 64} (1990) 1495, {\it Phys.

    Rev.} {\bf D44} (1991) 1369.

    A.V.Efremov, J.Soffer, O.V.Teryaev, {\it Nucl. Phys.} {\bf B346}
    (1990) 97.

    U.Ellwanger, B.Stech, {\it Phys. Lett. } {\bf B241} (1990) 409;
    {\it Z. Phys.} {\bf C49} (1991) 683.

    J.H.K\"uhn, V.I.Zakharov, {\it Phys. Lett.} {\bf B252} (1990) 615.\\
50. H.Abramowicz et al., {\it Z. Phys. } {\bf C 15} (1982) 19.\\
51. B.L.Ioffe, M.Karliner, {\it Phys. Lett.} {\bf 247B} (1990) 387.\\
52. B.V.Geshkenbein, B.L.Ioffe, {\it Nucl. Phys. } {\bf B166} (1980) 340.\\
53. D.Adams et al., {Preprint CERN-PPE/94-57} (1994).\\
54. R.M.Lombard-Nelson, {\it The Talk at ITEP Seminar} (unpublished).\\
55. B.Adeva et al., {\it Phys. Lett.} {\bf B302} (1993) 533.\\
56. P.L.Anthony et al., {\it Phys. Rev. Lett.} {\bf 71} (1993) 959.\\
57. J.Ellis, M.Karliner, {\it Phys. Lett.} {\bf B313} (1993) 131.\\
58. B.Adeva et al., {\it Phys. Lett.} {\bf B320} (1994) 400.\\
59. G.Altarelli, P.Nason, G.Ridolfi, {\it Phys. Lett. }  {\bf B320} (1994)
152.\\
60. V.D.Burkert, B.L.Ioffe, {\it Phys. Lett.} {\bf B296} (1992) 223.\\
61. V.D.Burkert, B.L.Ioffe, {\it ZhETF} {\bf 105} (1994) 1153.\\
62. V.D.Burkert, Zhujun Li, {\it Phys. Rev. } {\bf D47} (1993) 46.\\
63. E.Hughes, in {\it Lecture at SLAC Summer Institute on Particle Physics,}

    July 26 - August 6, 1993, Preprint SLAC-PUB-6439 (1994).\\
64. C.Y.Prescott, {\it Preprint SLAC-PUB-6428} (1994).\\
65. I.Karliner, {\it Phys. Rev.} {\bf D7} (1973) 2717.\\
66. R.L.Workman, R.A.Arndt, {\it Phys. Rev.}  {\bf D45} (1992) 1789.\\
67. D.Kaplan, A.Manohar, {\it Nucl. Phys.} {\bf B310} (1988) 527.\\
68. L.A.Ahrens et al., {\it Phys. Rev. } {\bf D35} (1987) 785.\\
69. B.L.Ioffe, {\it Phys. Lett.} {\bf 63B} (1976) 425.\\
70. S.Rabinowitz et al., {\it Phys. Rev. Lett. } {\bf 70} (1993) 134.

    A.O.Bazarko et al. (CCFR Collaboration), {\it Preprint Nevis} R1502
    (1994).\\
71. T.Bolton et al., {\it Phys. Lett. } {\bf B278} (1992) 495.\\
72. R.L.Jaffe, {\it Comm. Nucl. Part. Phys.}  {\bf 14} (1990) 239.\\
73. R.P.Feynman,  {\it Photon-Hadron Interaction} (W.A.Benjamen,
     New. York, 1971).\\
74. S.Wandzura, F.Wilczek, {\it Phys. Lett. } {\bf 72B} (1977) 195. \\
75. J.Ralston, D.E.Soper, {\it Nucl. Phys.} {\bf B152} (1979) 109.\\
76. R.L.Jaffe, X.Ji, {\it Phys. Rev. Lett. } {\bf 67} (1991) 552;

    {\it Nucl. Phys. }  {\bf B375} (1992) 527.\\
77. X.Artru, M.Mekhfi, {\it Z.Phys.} {\bf C45} (1990) 669.\\
78. J.C.Collins, {\it Nucl. Phys.} {\bf B394} (1993) 169.\\
79. B.L.Ioffe, A.Yu.Khodjamirian, {\it Munich Univ. Preprint}
    LMU-01/94 (1994);\\
80. R.L.Jaffe,
    in {\it Baryons'92}, Ed. by Moshe Gai (World Scientific, Singapore,
    1993),

    p.308.\\
81. B.L.Ioffe, {\it Pisma ZhETF } {\bf 42} (1985) 266,
    {\bf 43} (1986) 316.\\
82. V.M.Belyaev, B.L.Ioffe, {\it Nucl. Phys.} {\bf
    B310} (1988) 548.\\
83. J.Qui, G.Sterman, {\it Phys. Rev. Lett.} {\bf 67}
    (1991) 2264.\\
84. B.Carlitz et al., {\it Penn. State Univ. preprint }
    PSU/TH/101 (1992).\\
85. J.Collins, {\it Nuclear Phys.} {\bf B396} (1993)
    161.\\
86. A.V.Efremov, L.Mankiewicz, N.A.T\"ornquist, {\it Phys. Lett.}
    {\bf B284} (1992) 394.\\
\end{document}